%% file: ModulationAndCoding_Optnets_v05_arxiv.tex
\documentclass[journal,twoside,twocolumn,a4paper]{IEEEtran}

\usepackage{tikz,pgfplots}
\usetikzlibrary{positioning,shadows,shapes,fit,arrows,backgrounds}
\colorlet{figyellow}{yellow!40!white}
\colorlet{figred}{red!40!white}
\colorlet{figblue}{blue!20!white}
\colorlet{figgreen}{green!30!white}
\colorlet{figgray}{black!10!white}

\usepackage{amsmath,amssymb,float}
\usepackage{graphicx,multicol,multirow}
\usepackage{psfrag,float}
\newcommand{\tnr}[1]{{\textnormal{#1}}}
\newcommand{\ld}{\ldots}

\newcommand{\Rc}{R_{\tnr{c}}}

\newcommand{\OROM}{$1\Rc 1M$}
\newcommand{\TROM}{$2\Rc 1M$}
\newcommand{\ORVM}{$1\Rc$Var$M$}
\newcommand{\TRVM}{$2\Rc$Var$M$}

\newcommand{\SNRSevenPercent}{\tnr{SNR}_{7\%}}

\pgfplotsset{compat=1.12} 

\newcommand{\snrhorizontalbarsHD}[1]{
\pgfplotstableread{OH_HD_crossings_down.dat}{\firsttable}\pgfplotstablegetrowsof{OH_HD_crossings_down.dat}
\pgfplotstableread{OH_HD_crossings_down_shifted.dat}{\secondtable}\pgfplotstablegetrowsof{OH_HD_crossings_down_shifted.dat}
\pgfmathsetmacro{\rows}{\pgfplotsretval-1}
\foreach \i in {0,...,\rows}{%
  	\pgfplotstablegetelem{\i}{[index] 0}\of{\firsttable}\let\x\pgfplotsretval
  	\pgfplotstablegetelem{\i}{[index] 0}\of{\secondtable}\let\y\pgfplotsretval 
	\addplot+[thick,no marks] plot coordinates { (\x,#1) (\y,#1)};}
	}
\newcommand{\snrhorizontalbarsSD}[1]{
\pgfplotstableread{OH_SD_crossings_down.dat}{\firsttable}\pgfplotstablegetrowsof{OH_SD_crossings_down.dat}
\pgfplotstableread{OH_SD_crossings_down_shifted.dat}{\secondtable}\pgfplotstablegetrowsof{OH_SD_crossings_down_shifted.dat}
\pgfmathsetmacro{\rows}{\pgfplotsretval-1}
\foreach \i in {0,...,\rows}{%
  	\pgfplotstablegetelem{\i}{[index] 0}\of{\firsttable}\let\x\pgfplotsretval
  	\pgfplotstablegetelem{\i}{[index] 0}\of{\secondtable}\let\y\pgfplotsretval 
	\addplot+[thick,no marks] plot coordinates { (\x,#1) (\y,#1)};}
	}
\newcommand{\snrcrossingsHD}[2]{
\pgfplotstableread{OH_HD_crossings_down.dat}{\firsttable}\pgfplotstablegetrowsof{OH_HD_crossings_down.dat}
\pgfplotstableread{OH_SD_crossings_down.dat}{\secondtable}\pgfplotstablegetrowsof{OH_SD_crossings_down.dat}
\foreach \i in {0,...,\rows}{%
	\pgfplotstablegetelem{\i}{[index] 0}\of{\firsttable}\let\x\pgfplotsretval
	\addplot[color=black,thick,only marks,mark=asterisk] plot coordinates { (\x,#2)};
	\pgfplotstablegetelem{\i}{[index] 0}\of{\secondtable}\let\y\pgfplotsretval 
	\addplot[color=black,thick,only marks,mark=diamond*,mark size=1.5pt] plot coordinates { (\y,#1)};}
}
\def\innerradius{0.78cm}
\def\outerradius{1.3cm}
\newcommand{\wheelchart}[3]{
    \pgfmathsetmacro{\totalnum}{0}
    \foreach \value/\colour/\name in {#1} {
        \pgfmathparse{\value+\totalnum}
        \global\let\totalnum=\pgfmathresult
    }
    
\setlength{\fboxsep}{0.5pt}
\makebox[0.42\columnwidth]
{
    \begin{tikzpicture}[tight background] 

      \pgfmathsetmacro{\wheelwidth}{\outerradius-\innerradius}
      \pgfmathsetmacro{\midradius}{(\outerradius+\innerradius)/2}

      \begin{scope}[rotate=#2]

      \pgfmathsetmacro{\cumnum}{0}
      \foreach \value/\colour/\name in {#1} {
            \pgfmathsetmacro{\newcumnum}{\cumnum + \value/\totalnum*360}

            \pgfmathsetmacro{\percentage}{\value/\totalnum*100}
            \pgfmathsetmacro{\midangle}{-(\cumnum+\newcumnum)/2}

            \pgfmathparse{
               (-\midangle<180?"west":"east")
            } \edef\textanchor{\pgfmathresult}
            \pgfmathsetmacro\labelshiftdir{1-2*(-\midangle>180)}

            \fill[\colour] (-\cumnum:\outerradius) arc (-\cumnum:-(\newcumnum):\outerradius) --
            (-\newcumnum:\innerradius) arc (-\newcumnum:-(\cumnum):\innerradius) -- cycle;

            \draw  [*-,very thin] node [append after command={(\midangle:\midradius pt) -- (\midangle:\outerradius + 1.5ex) -- (\tikzlastnode)}] at (\midangle:\outerradius + 1.5ex) [xshift=\labelshiftdir*0.3cm,inner sep=0.2pt, outer sep=0.2pt,anchor=\textanchor]{\tiny{\name: \pgfmathprintnumber{\percentage}\%}};
	
            \global\let\cumnum=\newcumnum
        }

      \end{scope}
      \draw[black] (0,0) circle (\outerradius) circle (\innerradius);
	\node[draw=none,fill=none] {$#3$~Tbps};
    \end{tikzpicture}
    }
}

\usepackage[ddmmyyyy]{datetime}
\newcommand{\header}{Preprint, \today, \currenttime}
\title{On the Impact of Optimal Modulation and FEC Overhead on Future Optical Networks}
\author{Alex Alvarado, David J. Ives, Seb Savory and Polina Bayvel
\thanks{Research supported by Engineering and Physical Sciences Research Council (EPSRC) through the CDT in Photonic Systems Development and the project UNLOC (EP/J017582/1), and by the Royal Academy of Engineering/The Leverhulme Trust Senior Research Fellowship. This work was  presented in part at the 2015 Optical Fiber Communication Conference (OFC), Los Angeles, CA, Mar. 2015.}
\thanks{The authors are with the Optical Networks Group, Department of Electronic and Electrical Engineering, University College London, London WC1E~7JE, United Kingdom (email: alex.alvarado@ieee.org).}
}

\begin{document}


\maketitle

\begin{abstract}
The potential of optimum selection of modulation and forward error correction (FEC) overhead (OH) in future transparent nonlinear optical mesh networks is studied from an information theory perspective. Different network topologies are studied as well as both ideal soft-decision (SD) and hard-decision (HD) FEC based on demap-and-decode (bit-wise) receivers. When compared to the de-facto QPSK with 7\% OH, our results show large gains in network throughput. When compared to SD-FEC, HD-FEC is shown to cause network throughput losses of $12$\%, $15$\%, and $20$\% for a country, continental, and global network topology, respectively. Furthermore, it is shown that most of the theoretically possible gains can be achieved by using one modulation format and only two OHs. This is in contrast to the infinite number of OHs required in the ideal case. The obtained optimal OHs are between $5$\% and $80\%$, which highlights the potential advantage of using FEC with high OHs.
\end{abstract}


\section{Introduction and Motivation}

The rapid rise in the use of the Internet has led to increasing traffic demands putting severe pressure on backbone networks. The transport backbone of the Internet is formed of optical mesh networks, where optical fibre links connect nodes formed of reconfigurable optical add drop multiplexers (ROADMs). Studying the ultimate transmission limits of optical mesh networks as well as the optimal utilization of the installed network resources is therefore key to avoid the so-called ``capacity crunch'' \cite{ChralyvyECOC2009,Richardson2010}. 

Installed optical mesh networks utilize wavelength routing to transparently connect source and destination transceivers. The quality of the optical communication signal degrades due to transmission impairments, which in turn limits the maximum achievable data rate. This degradation is usually due to the amplifiers in the link as well as nonlinear distortion due to neighboring WDM channels. Furthermore, the optical signals within a transparent wavelength routed network travel a variety of distances, and thus, experience different levels of signal degradation. The most conservative design alternative for an optical network is to choose the transceiver to operate error-free on the worst light path, i.e., for transmission between the furthest spaced nodes \cite[p.~138]{Jinno2010}, \cite[Sec.~1]{Li14OSN}. Under this design paradigm, any route reconfiguration can be accommodated through the network. However, this leads to over provisioning of resources.

\begin{figure*}
\centerline{
\footnotesize{
\begin{tikzpicture}[>=stealth,auto,
block/.style={rectangle,rounded corners=3pt,thick,draw,inner sep=3pt,minimum width=22mm,
minimum height=10mm,fill=figyellow,drop shadow,align=center,execute at begin node=\setlength{\baselineskip}{2ex}},
plain/.style={align=center,execute at begin node=\setlength{\baselineskip}{2.5ex}}]
\newlength{\blocksep}\setlength{\blocksep}{8mm} 
\newlength{\vblocksep}\setlength{\vblocksep}{8mm} 
\draw[thick,fill=figgray,rounded corners=3pt,drop shadow] (-209pt,-23pt) rectangle (-50pt,25pt);
\node[plain,anchor=south west] at (-210pt,24pt) {CM Transmitter};
\draw[thick,fill=figgray,rounded corners=3pt,drop shadow] (78pt,8pt) rectangle (235pt,50pt);
\node[plain,anchor=south east] at (235pt,50pt) {HD BW Receiver};
\draw[thick,dashed,<->] (-128pt,13pt) -- (-128pt,70pt) -- (0pt,70pt) node[plain,above]{$1-H_{\tnr{b}}(\tnr{BER})$} -- (158pt,70pt) -- (158pt,43pt) ;
\draw[thick,fill=figgray,rounded corners=3pt,drop shadow] (78pt,-48pt) rectangle (235pt,-8pt);
\node[plain,anchor=south east] at (235pt,-8pt) {SD BW Receiver};
\draw[thick,dashed,->] (-128pt,50pt) -- (0pt,50pt) node[plain,above]{$\sum_{k=1}^{m} I(B_{k};L_{k})$} -- (65pt,50pt) -- (65pt,0pt) -- (158pt,0pt) -- (158pt,-16pt) ;
\node[block,fill=figred] (CHb) {Optical Channel};
\node[block,fill=figblue,left=\blocksep of CHb] (TXi) {$M$QAM Mapper\\with $M=2^{m}$};
\node[block,fill=figgreen,left=\blocksep  of TXi] (TXo) {FEC Encoder\\Rate $\Rc$};
\node[right=\blocksep of CHb] (RXi) {};
\node[above=\vblocksep of RXi] (RXiA) {};
\node[below=\vblocksep of RXi] (RXiB) {};
\node[block,fill=figblue,right= \blocksep of RXiA] (RXiAbove) {Hard-Decision\\$M$QAM\\Demapper};
\node[block,fill=figgreen,right=\blocksep of RXiAbove] (RXoAbove) {HD-FEC\\Decoder};
\node[block,fill=figblue,right= \blocksep of RXiB] (RXiBelow) {Soft-Decision\\$M$QAM\\Demapper};
\node[block,fill=figgreen,right=\blocksep of RXiBelow] (RXoBelow) {SD-FEC\\Decoder};
\draw[thick,->] (TXo) -- node[plain,above] {$\underline{B}$} (TXi);
\draw[thick,->] (TXi) -- node[plain,above]{$\underline{X}$} (CHb);
\draw[thick,dashed,<->] (-43pt,13pt) -- (-43pt,30pt) -- (0pt,30pt) node[plain,above]{$I(X;Y)$} -- (45pt,30pt) -- (45pt,13pt) ;
\draw[thick,-] (CHb) -- node[plain,above]{$\underline{Y}$} (RXi.center);
\draw[thick,-] (RXi.center) -- (RXiA.center);\draw[thick,->] (RXiA.center) -- (RXiAbove);
\draw[thick,-] (RXi.center) -- (RXiB.center);\draw[thick,->] (RXiB.center) -- (RXiBelow);
\draw[thick,->] (RXiAbove) -- node[plain,above]{$\hat{\underline{B}}$} node[plain,above,rotate=90,anchor=west] {} (RXoAbove);
\draw[thick,->] (RXiBelow) -- node[plain,above]{$\underline{L}$} node[plain,above,rotate=90,anchor=east] {} (RXoBelow);
\draw[->] (TXo.west) ++(-0.75\blocksep,0) -- node[plain,above]{$\underline{I}$} node[plain,rotate=90,anchor=east] {} (TXo);
\draw[->] (RXoAbove.east) -- node[plain,above]{$\hat{\underline{I}}$} node[plain,above,rotate=90,anchor=west] {} ++(0.75\blocksep,0);
\draw[->] (RXoBelow.east) -- node[plain,above]{$\hat{\underline{I}}$} node[plain,above,rotate=90,anchor=east] {} ++(0.75\blocksep,0);
\end{tikzpicture}}}
\caption{Coded modulation system under consideration. At the transmitter, a binary FEC code is concatenated with a $M$-ary QAM modulator. After transmission, the noisy received symbols are demapped and then decoded by a bit-wise receiver. When the demapper makes hard decisions on the symbols, an HD-FEC is used. When the demapper computes log-likelihood ratios (soft-decisions), an SD-FEC is assumed. The achievable rates discussed in Sec.~\ref{IT} are also shown.}
\label{model}
\end{figure*}
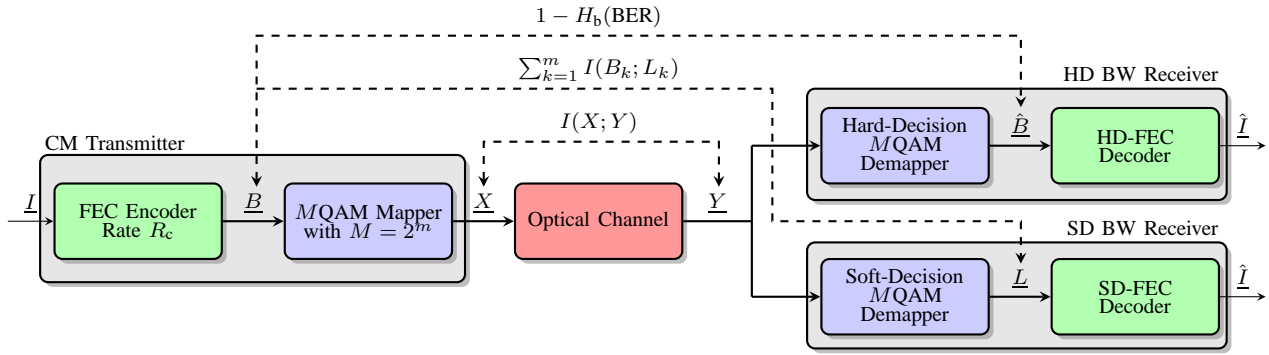

The increase in traffic demand together with the development of software-defined transceivers that can adapt the transmission parameters to the physical channel have increased the interest in designing networks that utilize the resources more efficiently. The degrees of freedom in the transceiver include, e.g., the forward error correction (FEC) scheme, FEC overhead (OH), modulation format, frequency separation (in flex-grid networks), launch power, and symbol rate (see \cite[Fig.~3]{Gringeri2013}). The network resources can be better utilized if these degrees of freedom are jointly optimized in conjunction with the routing of the optical light path through the network.

To cope with increasing capacity demands, future optical networks will use multilevel modulations and FEC. This combination is known as coded modulation (CM) and its design requires the joint optimization of the FEC and modulation format (see Fig.~\ref{model}). The optimum receiver structure for CM is the maximum likelihood (ML) receiver, which finds the most likely \emph{coded sequence} \cite[Sec.~3.1]{Alvarado15_Book}. The ML solution is in general impractical, and thus, very often the receiver is implemented as a (suboptimal) bit-wise (BW) receiver instead \cite[Sec.~3.2]{Alvarado15_Book}. In a BW receiver, hard or soft information on the code bits is calculated first, and then, an FEC decoder is used (see the receiver side of Fig.~\ref{model}). In other words, practical receivers decouple the detection process: symbols are first converted into bits, and then, FEC decoding is applied. In this paper we consider BW receivers with both soft-decision FEC (SD-FEC) and hard-decision FEC (HD-FEC). BW receivers have been studied for optical communications in, e.g., \cite{Bulow2011b,Millar14,Alvarado2015_JLT}

Traditional analyses of optical networks assume a target pre-FEC bit error rate (BER), and thus, an HD-FEC  with fixed OH is implicitly assumed.\footnote{When SD-FEC is considered,  however, pre-FEC BER does not determine the FEC OH, as recently shown in \cite{Alvarado2015b_JLT}.} Although the most common value for the OH is $7\%$, higher OHs have become increasingly popular. Furthermore, the use of SD-FEC with high OHs (typically around $20\%$) is considered the most promising FEC alternative for $100$G and $400$G transceivers. Although from a theoretical point of view, fixing the OH is an artificial constraint that reduces flexibility of the CM design and reduces the network throughput, there are good reasons for fixing the OH. The client rates are usually quantized to $10,40,100, 400$~Gb/s for compatibility with the Ethernet standards. The symbol rate is also often fixed to accommodate the transmitted bandwidth into a given fixed-grid, leading to fixed OHs. Under these constraints, no full flexibility on the selection of OHs is possible. In this paper, however, we will ignore these constraints and focus on finding the theoretically maximum network throughput.

To increase the network throughput, different approaches have been investigated in the literature. For example, \cite{Nag10} considered mixed line rates ($10,40,100$~Gb/s) for the NSF mesh topology, \cite{Gho11} considered variable FEC OHs with fixed symbol rate and modulation formats, and \cite{Gho12} considered variable modulation format and SD-FEC OHs. Adaptive FEC based on practical codes was recently considered in \cite{Li14OSN}. Variable OHs with $16$QAM and $64$QAM were studied in \cite{Mello14}, where probabilistically-shaped constellations were considered. The optimal modulation format based on an approximation for the maximum achievable rates of HD-FEC was considered in \cite{Savory14a}. Adaptive FEC OH for time frequency packing transmission was studied in \cite{Sambo2015OFC} and \cite{Sambo14}. The problem of routing and spectrum assignment for flex-grid optical networks with orthogonal frequency-division multiplexing was studied in \cite{Christodoulopoulos11}. The key enabling technology for these approaches are software-defined transceivers, allowing for example to vary the modulation format and symbol rate, as done in \cite{SchmogrowPTL2010,RobertsECOC2012,ChoiOpex2012,TeipenJON2012,ChoiPTL2013}. In \cite{Sambo2013OFC,Sambo14}, a software defined transceiver with variable FEC was experimentally demonstrated.

To optimize the network design, a physical layer model is required. While in the past very simple models (e.g., reach-based models) were considered, recently, nonlinear effects have been taken into account using the Gaussian noise (GN) model \cite{Splett93,Grellier11,Poggiolini11,beygi12,Johannisson13,Poggiolini14a,Johannisson14}. In \cite{Zhao15OFC}, the closed form solution of the GN model of \cite{Johannisson14} was used to adapt the routing and wavelength assignment problem, for a target pre-FEC BER, and 4 different modulation formats. The same GN model was used in \cite{Yan15OFC} and \cite{Yan15PTL} to jointly optimize power, modulation format, and carrier frequencies (flex-grid) for a fixed OH. Numerical integration of the GN model was used in \cite{Ives2014b} to assess SNR and throughput optimization via power and modulation adaptation. The numerically integrated GN model in \cite{Ives2014b} was also used in \cite{Ives2014a,Ives2015a} to sequentially optimize modulation format and power for a fixed FEC OH. The potential gains of adaptive FEC OH and modulation, or adaptive launch power and modulation were studied in \cite{Ives2015OFC}. 

In this paper, we study the problem of finding the optimal modulation and FEC OH from an information theory viewpoint. In particular, we use information theoretic quantities (i.e., achievable rates) and a realistic model for the nonlinear interference to study the maximum network throughput of optical mesh networks. For point-to-point links, and under a Gaussian assumption on the channel, the solution depends only on the signal-to-noise ratio (SNR). For an optical network, however, the solution depends on the SNR \emph{distribution} of the connections. Therefore, the theoretically optimum CM design is obtained when the modulation size and FEC OH are jointly designed \emph{across} the network. Significant increases in network throughput are shown. Furthermore, practically relevant schemes (based on either one or two OHs) are also considered, and their gap to the theoretical maximum is quantified. This paper extends our results in \cite{Alvarado15b} by considering multiple network topologies as well as both HD- and SD-FEC.

This paper is organized as follows. In Sec.~\ref{Prel} the system model, network topologies, and physical layer model are described. In Sec.~\ref{IT} the optimal selection of modulation and coding is reviewed and the maximum network throughput is analyzed. In Sec.~\ref{Practical} practically relevant schemes are considered. Conclusions are drawn in Sec.~\ref{Conc}.

\section{Preliminaries}\label{Prel}

\subsection{System Model}

We consider the CM transmitter shown in Fig.~\ref{model}, where a binary FEC code maps the \emph{information bits} $\underline{I}=[I_{1},I_{2},\ldots,I_{k_{\tnr{c}}}]$ into a sequence of code bits $\underline{B}=[B_{1},B_{2},\ldots,B_{n_{\tnr{c}}}]$, where $\Rc=k_{\tnr{c}}/n_{\tnr{c}}$ is the code rate. At each discrete-time instant, $m$ code bits are mapped to a constellation symbols from a discrete constellation with $M=2^m$ constellation points. We consider polarization-multiplexed square QAM constellations with $m=2,4,6,8,10$ (i.e., $M$QAM with $M\leq 1024$) in each polarization. The constellations are labeled by the binary-reflected Gray code. For FEC encoder with code rate $\Rc$ and a modulation format with $M$ symbols, the spectral efficiency (SE) per two polarizations is
\begin{align}\label{SE}
\tnr{SE}=2 \Rc m \left[\frac{\tnr{bit}}{\tnr{symbol}}\right].
\end{align}
The FEC OH is
\begin{align}\label{FEC.OH}
\tnr{OH}=100\left(\frac{1}{\Rc}-1\right)\%.
\end{align}

At the receiver side a bit-wise (BW) receiver is used\footnote{Also known as bit-interleaved coded modulation receiver \cite{Alvarado15_Book}}. In such a receiver, the noisy symbols are first demapped, and then, FEC decoding is performed. The FEC decoder gives an estimate of the transmitted bits $\hat{\underline{I}}$. Due to the separation of the detection process, BW receivers are suboptimal. They are, however, very popular in practice due to the use of off-the-shelf FEC decoders.

In this paper we consider two types of FEC decoders: hard- and soft-decision FEC. This naturally leads to two different BW receiver structures, shown on the right-hand side of Fig.~\ref{model}. In a hard-decision BW receiver, the demapper makes hard decision on the bits (by making hard-decision on the symbols), which are then passed to an HD-FEC decoder. We assume that there is a random bit-level interleaver between the encoder and mapper, which we consider part of the FEC encoder (decoder).

In a soft-decision BW receiver, the demapper calculates \emph{soft} information on the code bits (also known as ``soft bits''), which are then passed to an SD-FEC decoder. The soft information is usually represented in the form of logarithmic likelihood ratios (LLRs), defined as
\begin{align}\label{LLR.def}
L_{q}	& \triangleq \log\frac{f_{Y|B_{q}}(y|1)}{f_{Y|B_{q}}(y|0)},\, q=1,\ldots,m,
\end{align}
where $B_{q}$ is the $q$th bit at the input of the mapper, and $f_{Y|B_{q}}(y|b)$, $b\in\left\{0,1\right\}$ is the channel transition probability. The sign of the LLRs corresponds to a hard-decision on the bits, while its amplitude represents the reliability of the available information.

\subsection{Network Topologies}\label{Topology}

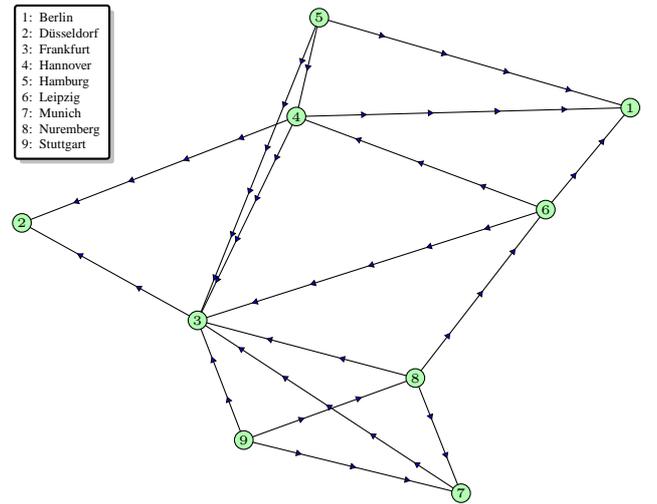
\begin{figure}[tbhp]
\centering{\footnotesize{
\begin{tikzpicture}
\input{DT2015.tikz}
\node[draw,text width=30pt,anchor=north west,thick,fill=white,rounded corners=1pt,drop shadow] at (-125pt,95pt) 
{\tiny{1: Berlin\\
2: D\"{u}sseldorf\\
3: Frankfurt\\
4: Hannover\\
5: Hamburg\\
6: Leipzig\\
7: Munich\\
8: Nuremberg\\
9: Stuttgart\\
}};
\end{tikzpicture}}}
\caption{DTG network formed by 9 nodes representative of a country-level topology. The location of the amplifiers are shown as triangles.}
\label{net:DTG}
\end{figure}

\begin{figure}[tbhp]
\centering{\footnotesize{
\begin{tikzpicture}
\input{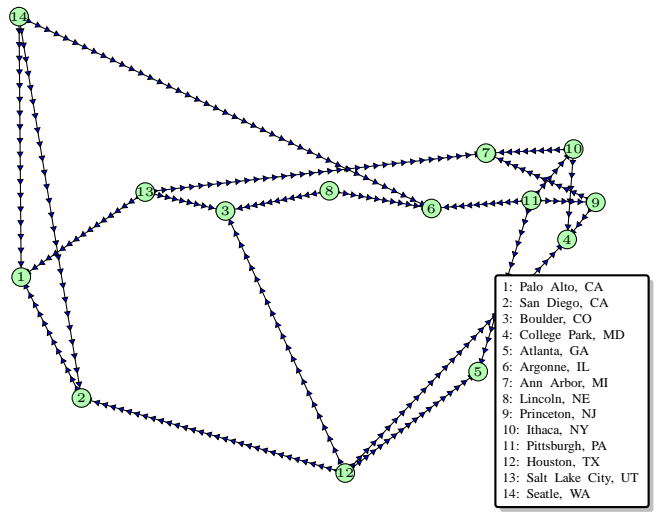}
\node[draw,text width=52pt,anchor=north east,thick,fill=white,rounded corners=1pt,drop shadow] at (114pt,-15pt) 
{\tiny{1: Palo Alto, CA\\
2: San Diego, CA\\
3: Boulder, CO\\
4: College Park, MD\\
5: Atlanta, GA\\
6: Argonne, IL\\
7: Ann Arbor, MI\\
8: Lincoln, NE\\
9: Princeton, NJ\\
10: Ithaca, NY\\
11: Pittsburgh, PA\\
12: Houston, TX\\
13: Salt Lake City, UT\\
14: Seatle, WA\\
}};
\end{tikzpicture}}}
\caption{NSF network formed by 14 nodes representative of a continental-level topology. The location of the amplifiers are shown as triangles.}
\label{net:NSFNET}
\end{figure}

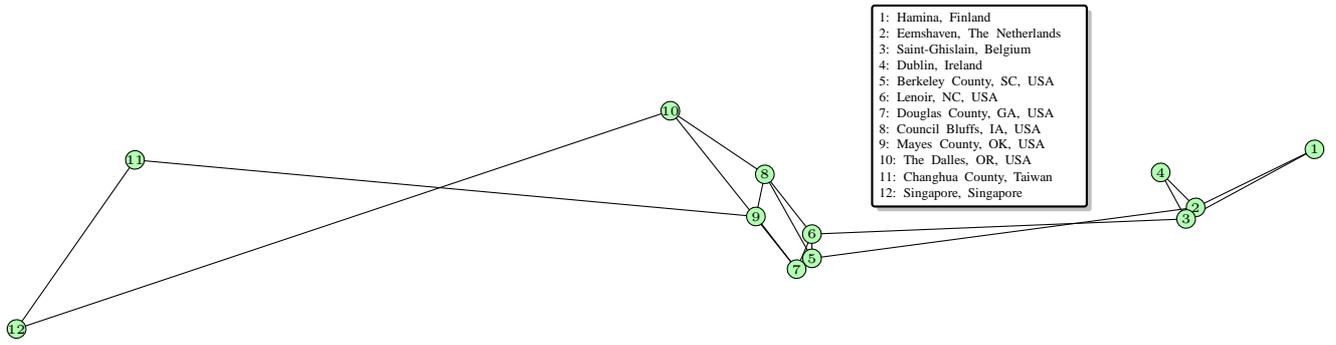
\begin{figure*}
\centering{\footnotesize{
\begin{tikzpicture}[rotate=-15]
\input{GoogleB4.tikz}
\node[draw,text width=75pt,anchor=south west,thick,fill=white,rounded corners=1pt,drop shadow] at (25pt,7pt) 
{\tiny{1: Hamina, Finland\\
		2: Eemshaven, The Netherlands\\
		3: Saint-Ghislain, Belgium\\
		4: Dublin, Ireland\\
		5: Berkeley County, SC, USA\\
		6: Lenoir, NC, USA\\
		7: Douglas County, GA, USA\\
		8: Council Bluffs, IA, USA\\
		9: Mayes County, OK, USA\\
		10: The Dalles, OR, USA\\
		11: Changhua County, Taiwan \\
		12: Singapore, Singapore\\
	}};
\end{tikzpicture}}}
\caption{GB4 network (amplifiers are not shown) formed by 12 nodes representative of a global-level topology.}
\label{net:GoogleB4}
\end{figure*}

In this paper we consider 3 networks shown in Figs.~\ref{net:DTG}, \ref{net:NSFNET}, and \ref{net:GoogleB4}. The first one is the exemplary network topology for Deutsche Telekom Germany (DTG) \cite[Sec.~II]{Makovejs2015JOCN}, where the two core nodes per city (see \cite[Fig.~1]{Makovejs2015JOCN}) were merged into one. The second one is the reference $14$-node, $21$-link NSF mesh topology \cite[Fig.~1]{Ives2014a}, \cite[Fig.~7]{Ives2014b}. The last topology is the Google B4 (GB4) network connecting data centers in \cite[Fig.~1]{Jain2013}. We chose to study these three topologies because they are representative of networks at three different scales: country, continental, and global.

\subsection{Physical Layer Model}\label{PHY}

For the analysis in this paper, it is assumed that each node in the networks described in Sec.~\ref{Topology} is equipped with multiple transceivers. Furthermore, it is also assumed that these transceivers are based on the structure in Fig.~\ref{model} and that they can ideally adapt the FEC OH and modulation order.  We consider a fixed grid of $80$ WDM channels of $50$~\tnr{GHz} and a symbol rate of $32$~\tnr{GBaud}. 
 
The nodes are connected with fiber pairs with standard single-mode fiber with parameters shown in Table~\ref{tab:constants}. Erbium-doped fiber amplifiers (EDFA) are regularly placed between the links, as shown in Figs.~\ref{net:DTG} and \ref{net:NSFNET}\footnote{The amplifiers are not shown in Fig.~\ref{net:GoogleB4} so as not to overcrowd the figure.}. The span length is $80$~\tnr{km} and the EDFA noise figure is $5$~\tnr{dB}.

Every link in the network is modeled using a channel model (see Fig.~\ref{model}) that encompasses all the transmitter digital signal processing (DSP) used after the mapper (i.e., pulse shaping, polarization multiplexing, filtering, electro-optical conversion, etc.), the physical channel (the fiber and amplifiers), and the receiver DSP (optical-to-electrical conversion, filtering, equalization, matched filtering, etc.). This channel is modeled using a dual-polarization, discrete-time, memoryless, additive white Gaussian noise (AWGN) channel. 

For each polarization, and at each discrete time $k=1,2,\ldots,n$, $Y_{k}=X_{k}+Z_{k}$, where $X_{k}$ is the transmitted symbol, $Z_{k}$ are independent, zero-mean, circularly symmetric, complex Gaussian random variables, and $n$ is the blocklength. This GN channel model characterizes optically-amplified links dominated by amplified spontaneous emission noise where chromatic dispersion and polarization mode dispersion are perfectly compensated. In this model, the power-dependent nonlinearities in the presence of sufficient dispersion are treated as an additional source of AWGN.

The symbol SNR for a route with $N_{\tnr{s}}$ spans is calculated as
\begin{align}\label{SNR.GN}
\tnr{SNR} = \frac{P}{N_{\tnr{s}}(P_{\tnr{ASE}}+\eta P^{3})}
\end{align}
where $P$ is the launch power per channel, $P_{\tnr{ASE}}$ is the ASE noise added after each span (in the $32$~GHz signal bandwidth), $\eta$ is the nonlinear coefficient (per span).

The nonlinear interference is taken as that on the worst case central DWDM channel, i.e., we assumed that all the links were fully loaded with DWDM channels. The nonlinear coefficient $\eta$ is calculated using the incoherent GN model of nonlinear interference \cite{Poggiolini14a}, SPM is assumed to be ideally compensated, and the ROADM nodes were assumed lossless. Using the parameters in Table~\ref{tab:constants}, we obtain $\eta \approx 742~\tnr{W}^{-2}$. The launch power that maximizes the SNR in \eqref{SNR.GN} is found to be $-1$~\tnr{dBm}. A summary of the parameters discussed in this section is given in Table~\ref{tab:constants}.

\begin{table}[tpb]
	\renewcommand{\arraystretch}{1.2}
	\caption{Summary of system parameters.}
	\centering
	\begin{tabular}{ l | l }
\hline
	Parameter		 				& Value \\
\hline 

\hline
		Fiber attenuation				& $0.22~\tnr{dB/km}$ 			\\
		Dispersion parameter			& $16.7~\tnr{ps}/\tnr{nm}/\tnr{km}$	\\
		Fiber nonlinear coefficient			& $1.3~\tnr{1/W/km}$ 			\\
		Span length					& $80$~\tnr{km}				\\
		EDFA noise figure 				& $5$~\tnr{dB}					\\
		ASE Noise per span	$P_{\tnr{ASE}}$& $0.747~\mu\tnr{W}$			\\
\hline

\hline
		Symbol rate				& $32~\tnr{Gbaud}$				\\
		WDM channels				& $80$						\\
		Channel separation			& $50$~\tnr{GHz}				\\
		Pulse shape				& Nyquist sinc pulses			\\
		SPM 					& Ideally compensated\\
		ROADM Loss				& $0$~\tnr{dB}\\
\hline

\hline
		Nonlinear coefficient $\eta$	& $742~\tnr{1/W$^2$}$			\\
		Launch power per channel $P$	& $-1$~\tnr{dBm}				\\
\hline

\hline
	\end{tabular}
	\label{tab:constants}
\end{table}


\subsection{Performance Metric}

Throughout this paper, the main performance metric considered is the total \emph{network throughput}, which we define below and denote by $\Theta$. The network throughput is the total traffic transported by the network that satisfies the required traffic profile.

The network is assumed to have $N$ nodes. The required connectivity is defined by the normalized traffic profile $T_{s,d}$ with $s,d\in\mathcal{N}$, where $\mathcal{N}\triangleq\{1,\ldots,N\}$ is the set of network nodes. The traffic profile is normalized such that
\begin{align}
\sum_{s=1}^{N} \sum_{d=1}^{N} T_{s,d} = 1.
\end{align}
We assume connectivity between all pairs of nodes is required ($0<T_{s,d}<1, s\neq d$) and that $T_{s,d}=0$ if $s=d$.

Let $C_{s,d}$ be the total available throughput between nodes $s$ and $d$ (across different routes), where
\begin{align}\label{Th.sd}
C_{s,d} \triangleq  \sum_{r=1}^{R_{s,d}} C_{s,d}^{(r)}
\end{align}
where $C_{s,d}^{(r)}$ is the available throughput in the $r$th route, and $R_{s,d}$ is the number of active routes between nodes $s$ and $d$. The network throughput $\Theta$ is then defined as
\begin{align}\label{Th.def}
\Theta \triangleq  \min_{\substack{s,d \in \mathcal{N}\\ d\neq s}}  \frac{C_{s,d}}{T_{s,d}}.
\end{align}
In this paper, we consider a uniform traffic profile, i.e.,
\begin{align}\label{T.uniform}
T_{s,d} = &
\begin{cases}
\frac{1}{N (N-1)}, 		& s \neq d \\
0, 					&  s = d
\end{cases}.
\end{align}
The network throughput in this case can be expressed as
\begin{align}\label{Th.def.uniform}
\Theta &= N(N-1) \min_{\substack{s,d \in \mathcal{N}\\ d\neq s}} C_{s,d} \\
\label{Th.uniform.final}
	& = N(N-1) \min_{\substack{s,d \in \mathcal{N}\\ d\neq s}} \sum_{r=1}^{R_{s,d}} C_{s,d}^{(r)}
\end{align}
where \eqref{Th.uniform.final} follows from \eqref{Th.sd}.

\section{Optimal Modulation and FEC Overhead}\label{IT}

In this section, we describe the selection of optimal FEC OH and modulation format from an information theory viewpoint. We first consider the ideal case of continuous constellations and then move to the case of modulation with discrete number of constellation points. The routing and wavelength assignment problem and the maximum network throughput are also discussed in this section.

\subsection{Channel Capacity}\label{IT:RatesAWGN}

The capacity of the AWGN channel (in [bit/sym]) under an average power constraint is 
\begin{align}\label{C.AWGN}
C=2\log_{2}(1+\tnr{SNR})
\end{align}
where the pre-log factor of 2 takes into account the two polarizations. The value of $C$ represents the maximum number of information bits per symbol that can be reliably transmitted through an AWGN channel. 

The capacity in \eqref{C.AWGN} is achieved when the transmitted symbols are chosen from a zero-mean Gaussian distribution. In practice, however, the modulation has $M$ discrete levels, which reduces the achievable transmission rates. This case is studied in the next section.

\subsection{Achievable Rates for Discrete Constellations}\label{IT:RatesDCs}

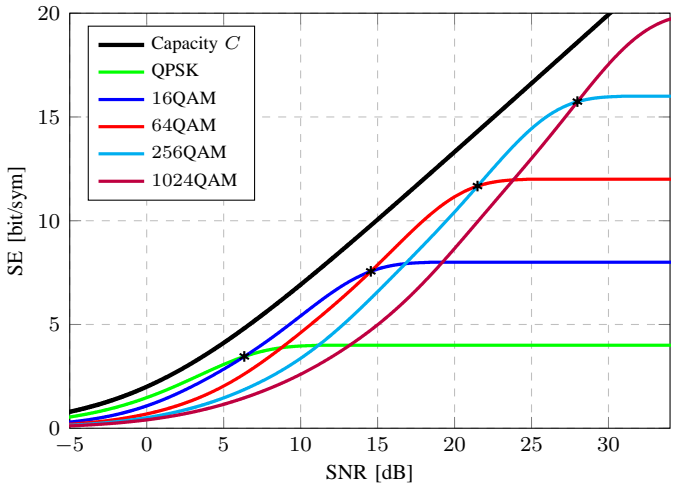
\begin{figure}[tbp]
\centering
\begin{tikzpicture}
\pgfplotscreateplotcyclelist{color list}{black,green,blue,red,cyan,purple,yellow}
\begin{axis}[cycle list name=color list,
	xminorgrids=true,
        width=1.07\columnwidth,
        height=0.8\columnwidth,
        grid=both,
        xmin=-5,xmax=34,
        ymin=0,ymax=20,
        xlabel={$\tnr{SNR}$~[dB]},
        xlabel style={yshift=0.1cm},
        ylabel={SE~[bit/sym]},
        ylabel style={yshift=-0.1cm},
        xtick={-5,0,...,30},
        ytick={0,5,...,20},
        every axis/.append style={font=\footnotesize},
	legend entries={Capacity $C$,QPSK,$16$QAM,$64$QAM,$256$QAM,$1024$QAM},
	legend style={legend pos=north west,font=\scriptsize,legend cell align=left},
	grid style={dashed}
	]
\addplot+[ultra thick] file {AWGN_HD.dat};
\foreach \i in {1,2,...,5} {\addplot +[very thick] file {MI_HD_m_\i.dat}; }%
\addplot[color=black,thick,only marks,mark=asterisk] file {MI_HD_crossings.dat};
\end{axis}
\end{tikzpicture}
\caption{Achievable rates for a BW receiver with HD-FEC. The black asterisks show the SNR values where the modulation size should be changed. The channel capacity in \eqref{C.AWGN} is shown for comparison.}
\label{MI_vs_SNR_HD}
\end{figure}

From an information theory point of view, the optimal code rate and constellation size can be chosen from the \emph{mutual information} (MI). The MI, usually denoted by $I(X;Y)$, is an achievable rate for an optimal receiver.\footnote{In information theory, a transmission rate $R$ is said to be achievable if there exist an encoder with rate $R$ and a (possibly very complex) decoder, such that the error probability after decoding vanishes as $n\to \infty$.} MI curves for square QAM constellations indicate that, regardless of the SNR, in order to maximize the SE, the densest available constellation should always be used and the code rate chosen between $0$ and $1$.\footnote{In practice, of course, this is difficult as using very dense constellations at low SNRs are hard to realize.} This has been shown, .e.g., in \cite[Fig.~4.3]{Alvarado15_Book}, \cite[Fig.~1]{Mello14}, \cite[Fig.~11]{Poggiolini14a}.

The MI is not an achievable rate for the two receiver structures we consider in this paper (see Fig.~\ref{model}). The first receiver in Fig.~\ref{model} is suboptimal because it makes hard-decisions on the symbols (and thus, information is lost). The second receiver is suboptimal because the LLR calculation ignores the dependency of the bits within a symbol (i.e., $L_{q}$ in \eqref{LLR.def} does not depend on $B_{l}$ for $l\neq q$). In this paper we consider two different achievable rates, one for each of these receiver structures.

For the case of a BW receiver with hard-decisions, Shannon's coding theorem state that error-free transmission is possible when $n\rightarrow\infty$ if the rate of the encoder fulfills \cite[eq.~(5)]{Savory14a} \cite[eq.~(8)]{Fehenberger15a}
\begin{align}
\Rc	& \leq \frac{I(B;\hat{B})}{m}
\label{C.DIDO}
	=1-H_{\tnr{b}}(\tnr{BER})
\end{align}
where $I(B;\hat{B})$ is the MI between the transmitted and received code bits (see Fig.~\ref{model}). In \eqref{C.DIDO}, $H_{\tnr{b}}(p)=-p\log_{2}{p}-(1-p)\log_{2}{(1-p)}$ is the binary entropy function and $\tnr{BER}$ is the average pre-FEC BER (across $m$ bit positions).

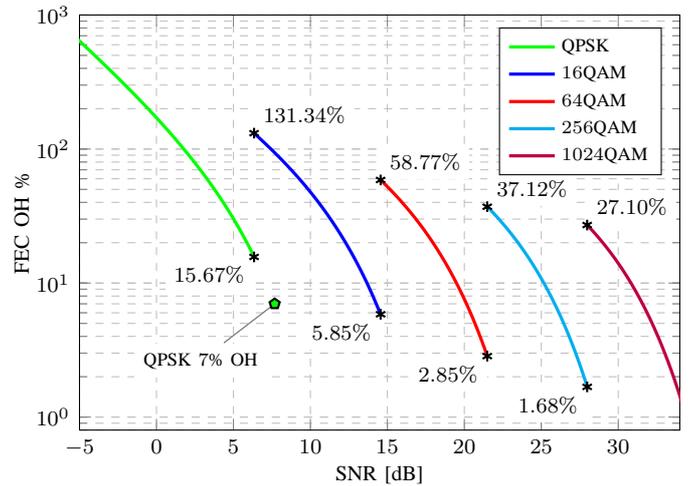
\begin{figure}[tbp]
\tikzstyle{every pin}=[fill=white,font=\scriptsize]
\centering
\begin{tikzpicture}
\pgfplotscreateplotcyclelist{color list}{green,blue,red,cyan,purple,yellow}
\begin{semilogyaxis}[cycle list name=color list,
	xminorgrids=true,
        width=1.07\columnwidth,
        height=0.8\columnwidth,
        grid=both,
        xmin=-5,xmax=34,
        ymin=.8,ymax=1000,
        xlabel={$\tnr{SNR}$~[dB]},
        xlabel style={yshift=0.1cm},
        ylabel={FEC OH~\%},
        ylabel style={yshift=-0.15cm},
        xtick={-5,0,...,30},
        every axis/.append style={font=\footnotesize},
	legend entries={QPSK,$16$QAM,$64$QAM,$256$QAM,$1024$QAM},
	legend style={legend pos=north east,font=\scriptsize,legend cell align=left},
	grid style={dashed},
	]
\foreach \i in {1,2,...,5} {\addplot +[very thick] file {OH_HD_m_\i.dat}; }%
\pgfplotstableread{OH_HD_crossings_down.dat}{\firsttable}   
\pgfplotstablegetrowsof{OH_HD_crossings_down.dat}
\pgfmathsetmacro{\rows}{\pgfplotsretval-2}      
\foreach \i in {1,...,\rows}{%
  	\pgfplotstablegetelem{\i}{[index] 0}\of{\firsttable} 
	\let\x\pgfplotsretval 
  	\pgfplotstablegetelem{\i}{[index] 1}\of{\firsttable} 
	\let\y\pgfplotsretval 
	\addplot[color=black,thick,only marks,mark=asterisk] plot coordinates { (\x,\y) } ;
	\edef\temp{\noexpand\node[fill=white,draw=none,anchor=north east] at (axis cs:\x,\y) {$\y \%$};}
	\temp}
\pgfplotstableread{OH_HD_crossings_up.dat}{\firsttable}   
\pgfplotstablegetrowsof{OH_HD_crossings_up.dat}
\pgfmathsetmacro{\rows}{\pgfplotsretval-1}      
\foreach \i in {0,...,\rows}{%
  	\pgfplotstablegetelem{\i}{[index] 0}\of{\firsttable} 
	\let\x\pgfplotsretval 
  	\pgfplotstablegetelem{\i}{[index] 1}\of{\firsttable} 
	\let\y\pgfplotsretval 
	\addplot[color=black,thick,only marks,mark=asterisk] plot coordinates { (\x,\y) } ;
	\edef\temp{\noexpand\node[fill=white,draw=none,anchor=south west] at (axis cs:\x,\y) {$\y \%$};}
	\temp}
	\addplot[color=black,mark=pentagon*,fill=green,thick,only marks,mark size=2pt] coordinates {(7.68,7)} node[pin=260:{QPSK $7$\% OH},inner sep=1pt]{};
\end{semilogyaxis}
\end{tikzpicture}
\caption{OHs obtained from the achievable rates in Fig.~\ref{MI_vs_SNR_HD} for a BW receiver with HD-FEC. The black asterisks show the SNR values where the modulation size should be changed and the corresponding FEC OH. The channel capacity in \eqref{C.AWGN} is shown for comparison. The SNR required for QPSK with $7$\% FEC OH is also shown with a green pentagon.}
\label{OH_vs_SNR_HD}
\end{figure}
  
The values of $I(B;\hat{B})$ for $M=4,16,\ld,1024$ are shown in Fig.~\ref{MI_vs_SNR_HD}, where the BER was calculated using \cite[Th.~2]{Ivanov12a}. The key difference between the achievable rates $I(B;\hat{B})$ and $I(X;Y)$ is that the former cross each other for different values of $M$ (see the asterisks in Fig.~\ref{MI_vs_SNR_HD}). An important consequence of crossing achievable rate curves is that the theoretically optimal choice of $\Rc$ and $M$ is not straightforward. For a BW receiver with HD-FEC we consider here, QPSK should be used for SNRs below $\tnr{SNR}\leq 5.8$~dB, $16$QAM for $5.8\leq \tnr{SNR}\leq 14$~dB, etc. The corresponding FEC OHs obtained via \eqref{FEC.OH} are shown in Fig.~\ref{OH_vs_SNR_SD}. This figure also shows the optimum minimum and maximum OH values for each modulation format as well as the SNR required for QPSK with $7$\% FEC OH (green pentagon).

When the BW receiver operates with soft-decisions, and if the code bits are independent, an achievable rate is given by the GMI \cite[eq.~(4.55)]{Alvarado15_Book}, \cite[eq.~(13)]{Alvarado2015_JLT}, \cite[eq.~(25)]{Alvarado2015b_JLT}
\begin{align}\label{GMI.def}
\tnr{GMI} = \sum_{k=1}^{m} I(B_{k};Y)=\sum_{k=1}^{m} I(B_{k};L_{k})
\end{align}
where the second equality holds if the LLRs are calculated via \eqref{LLR.def} and where $I(B_{k};L_{k})$ is the MI between the code bits and LLRs before FEC decoding. The three achievable rates considered above (MI, \eqref{C.DIDO} and \eqref{GMI.def}) are schematically shown in Fig.~\ref{model}.

\begin{figure}[!tbp]
\centering
\begin{tikzpicture}
\pgfplotscreateplotcyclelist{color list}{black,green,blue,red,cyan,purple,yellow}
\begin{axis}[cycle list name=color list,
	xminorgrids=true,
        width=1.07\columnwidth,
        height=0.78\columnwidth,
        grid=both,
        xmin=-5,xmax=29,
        ymin=0,ymax=20,
        xlabel={$\tnr{SNR}$~[dB]},
        xlabel style={yshift=0.1cm},
        ylabel={SE~[bit/sym]},
        ylabel style={yshift=-0.1cm},
        xtick={-5,0,...,30},
        ytick={0,5,...,20},
        every axis/.append style={font=\footnotesize},
	legend entries={Capacity $C$,QPSK,$16$QAM,$64$QAM,$256$QAM,$1024$QAM},
	legend style={legend pos=north west,font=\scriptsize,legend cell align=left},
	grid style={dashed}
	]
\addplot+[ultra thick] file {AWGN_SD.dat};
\foreach \i in {1,2,...,5} {\addplot +[very thick] file {MI_SD_m_\i.dat}; }%
\addplot[color=black,thick,only marks,mark=diamond*] file {MI_SD_crossings.dat};
\addplot[color=black,thick,only marks,mark=asterisk] file {MI_HD_crossings.dat};
\end{axis}
\end{tikzpicture}
\caption{Achievable rates (GMI) for a BW receiver with SD-FEC. The black diamonds show the SNR values where the modulation size should be changed. The channel capacity in \eqref{C.AWGN} and the crossing points of the achievable rates for HD-FEC (asterisks) taken from Fig.~\ref{MI_vs_SNR_HD} are also shown for comparison.}
\label{MI_vs_SNR_SD}
\end{figure}
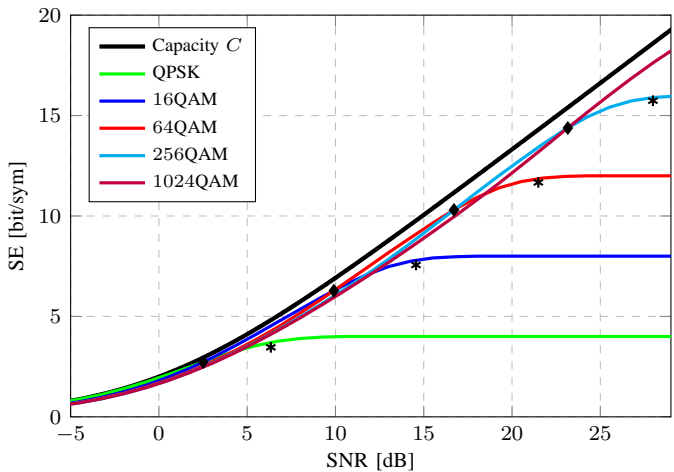

Fig.~\ref{MI_vs_SNR_SD} shows the GMI in \eqref{GMI.def} for different constellations. Similarly to the achievable rates for HD-FEC, the GMI curves cross each other (see black diamonds).\footnote{We again emphasize that this is only due to the fact a suboptimal receiver is considered. MI curves, on the other hand, do not cross each other for $M$QAM constellations, which has been known for many years (see, e.g., \cite[Fig.~2]{Ungerboeck82}, \cite[Fig.~1]{Forney98}).} Although this effect is less noticeable, the theoretical implications are the same: different SNRs call for different modulation sizes and FEC OH. In Fig.~\ref{MI_vs_SNR_SD}, we also show (with asterisks) the crossing points of the achievable rates for HD-FEC taken from Fig.~\ref{MI_vs_SNR_HD}. We do this to emphasize that if SD-FEC is considered instead of HD-FEC, different (higher) SNR thresholds are obtained. This is also visible in Fig.~\ref{OH_vs_SNR_SD}, where the optimum FEC OHs for SD-FEC are shown. The results in Figs.~\ref{MI_vs_SNR_SD} and \ref{OH_vs_SNR_SD} have been recently experimentally studied in \cite{MaherSPPCOM2015,MaherECOC2015}.

\subsection{Routing and Wavelength Assignment Problem}\label{NLIP}

The routing and wavelength assignment problem is solved numerically as an integer linear programming (ILP) problem as described in \cite[Section 4.1]{Ives2015a}. In particular, we maximize the network throughput in \eqref{Th.uniform.final} (i.e., under a uniform traffic demand), where $C_{s,d}$ is assumed to be given by the capacity function $C$ in \eqref{C.AWGN}. The ILP solution provides the number of active light paths and their routes between each node pair. From this, the total number of active light paths in the network is obtained. This solution also provides the SNR of each active light path.

\begin{figure}[!tbp]
\tikzstyle{every pin}=[fill=white,font=\scriptsize]
\centering
\begin{tikzpicture}
\pgfplotscreateplotcyclelist{color list}{green,blue,red,cyan,purple,yellow}
\begin{semilogyaxis}[cycle list name=color list,
	xminorgrids=true,
        width=1.07\columnwidth,
        height=0.78\columnwidth,
        grid=both,
        xmin=-5,xmax=29,
        ymin=6,ymax=600,
        xlabel={$\tnr{SNR}$~[dB]},
        xlabel style={yshift=0.1cm},
        ylabel={FEC OH~\%},
        ylabel style={yshift=-0.15cm},
        xtick={-5,0,...,30},
        every axis/.append style={font=\footnotesize},
	legend entries={QPSK,$16$QAM,$64$QAM,$256$QAM,$1024$QAM},
	legend style={legend pos=north east,font=\scriptsize,legend cell align=left},
	grid style={dashed}
	]
\foreach \i in {1,2,...,5} {\addplot +[very thick] file {OH_SD_m_\i.dat}; }%
\pgfplotstableread{OH_SD_crossings_down.dat}{\firsttable}   
\pgfplotstablegetrowsof{OH_SD_crossings_down.dat}
\pgfmathsetmacro{\rows}{\pgfplotsretval-2}      
\foreach \i in {1,...,\rows}{%
  	\pgfplotstablegetelem{\i}{[index] 0}\of{\firsttable} 
	\let\x\pgfplotsretval 
  	\pgfplotstablegetelem{\i}{[index] 1}\of{\firsttable} 
	\let\y\pgfplotsretval 
	\addplot[color=black,thick,only marks,mark=diamond*] plot coordinates { (\x,\y) } ;
	\edef\temp{\noexpand\node[fill=white,draw=none,anchor=north east] at (axis cs:\x,\y) {$\y \%$};}
	\temp}
\pgfplotstableread{OH_SD_crossings_up.dat}{\firsttable}   
\pgfplotstablegetrowsof{OH_SD_crossings_up.dat}
\pgfmathsetmacro{\rows}{\pgfplotsretval-1}      
\foreach \i in {0,...,\rows}{%
  	\pgfplotstablegetelem{\i}{[index] 0}\of{\firsttable} 
	\let\x\pgfplotsretval 
  	\pgfplotstablegetelem{\i}{[index] 1}\of{\firsttable} 
	\let\y\pgfplotsretval 
	\addplot[color=black,thick,only marks,mark=diamond*] plot coordinates { (\x,\y) } ;
	\edef\temp{\noexpand\node[fill=white,draw=none,anchor=south west] at (axis cs:\x,\y) {$\y \%$};}
	\temp}
	\addplot[color=black,mark=pentagon*,fill=green,thick,only marks,mark size=2pt] coordinates {(6.56,7)} node[pin=135:{QPSK $7$\% OH},inner sep=1pt]{};
\end{semilogyaxis}
\end{tikzpicture}
\caption{OHs obtained from the achievable rates in Fig.~\ref{MI_vs_SNR_SD} for a BW receiver with SD-FEC. The black diamonds show the SNR values where the modulation size should be changed and the corresponding FEC OH. The channel capacity in \eqref{C.AWGN} is shown for comparison.The SNR required for QPSK with $7$\% FEC OH ($\SNRSevenPercent$) is also shown with a green pentagon.}
\label{OH_vs_SNR_SD}
\end{figure}

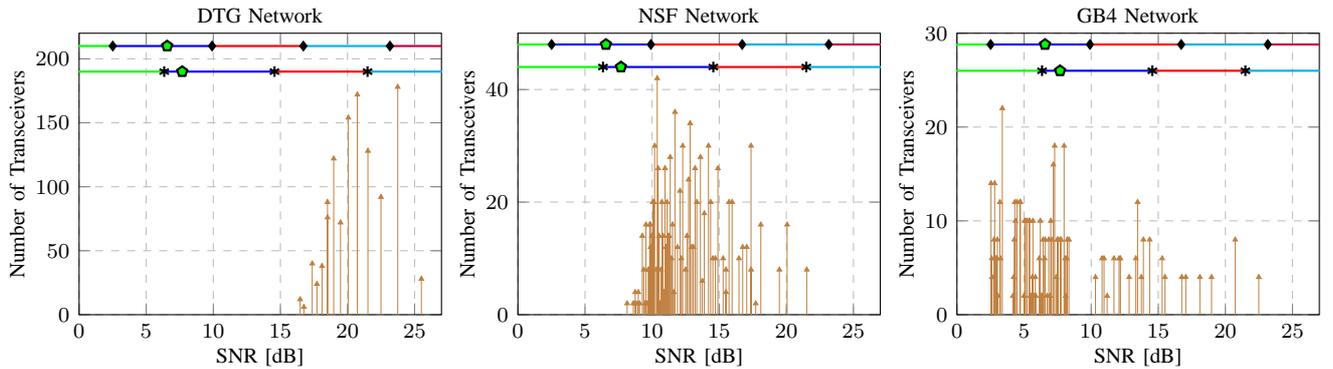
\begin{figure*}
\begin{center}
\input{snr_results.tex}
\end{center}
\caption{Number of transceivers (vertical lines with triangles at the top) for all SNR values for the three network topologies in Figs.~\ref{net:DTG}, \ref{net:NSFNET}, and \ref{net:GoogleB4}. These values are obtained by assuming all transceivers can achieve the capacity $C$ in \eqref{C.AWGN}. The horizontal lines show the SNR ranges in which different formats should be used. As in Figs.~\ref{OH_vs_SNR_HD} and \ref{OH_vs_SNR_SD}, asterisks and diamonds represent the SNR thresholds for HD- and SD-FEC, resp. The SNR thresholds for QPSK with $7$\% OH are also shown with green pentagons.}
\label{snr_results}
\end{figure*}

The SNR values obtained by solving the ILP problem for the three network topologies in Figs.~\ref{net:DTG}, \ref{net:NSFNET}, and \ref{net:GoogleB4} are shown in Fig.~\ref{snr_results}. The vertical bars show the number of transceivers that need to be installed to maximize the network throughput. The number of (two-way) transceivers for the DTG, NSF, and DTG networks are $1230$, $1094$, and $570$, resp.

The vertical bars in Fig.~\ref{snr_results} can be interpreted as the \emph{distribution} of SNR across the network. By comparing these distributions, it is clear that the average SNRs across the network decreases as the size of the network increases. This is due to the presence of long links in continental and global networks (NSF and GB4). The SNR distributions in Fig.~\ref{snr_results} also show that the \emph{spread} of the SNR values is much larger for large networks. While for the DTG network, the variation in SNR is about $10$~dB, for the GB4 network, this variation is about $20$~dB.

Once the SNR values for the active light paths are found, the maximum network throughput can be calculated via \eqref{Th.uniform.final}. In particular, the SNRs of the routes shown in Fig.~\ref{snr_results} are first grouped for each source destination pair. Then, the SNRs are ``mapped'' to throughputs via \eqref{C.AWGN}, and the value of $ \sum_{r=1}^{R_{s,d}} C_{s,d}^{(r)}$ in \eqref{Th.uniform.final} is obtained. The resulting throughputs are $524$, $278$, and $88$~Tbps, for the DTG, NSF, and GB4 networks, resp. These throughput values are shown in the first columns of Table~~\ref{tab:ideal} together with the number of transceivers for each network.

\begin{table*}
	\renewcommand{\arraystretch}{1.2}
	\caption{Optimum number of transceivers and network throughput $\Theta$ (in Tbps) under idealistic assumptions on coding and modulation.}
	\centering
	\begin{tabular}{cccccc}
\hline

\hline
					& & &  \multicolumn{3}{c}{Network Throughput $\Theta$ (in [Tbps])} \\
\hline

\hline
		Network 	& Nodes  $N$& Transceivers	& Maximum (Capacity-based)	& Ideal HD-FEC	& Ideal SD-FEC \\
\hline 

\hline
		DTG			& $9$ & $1230$		& $524$		& $431$ 		& 	$488$\\
		NSFNET		& $14$& $1094$		& $278$		& $217$ 		& 	$255$\\
		GB4			& $12$ & $570$		& $88$		& $64$ 			& 	$81$\\
\hline

\hline
	\end{tabular}
	\label{tab:ideal}
\end{table*}

\subsection{Ideal FEC}\label{IdealFEC}

The results in the previous section assume all the transceivers can achieve the capacity of the the AWGN channel. This is never the case in practice as it requires the use of continuous constellations. In this section we  we consider the case where all transceivers can choose any of the $M$QAM constellations considered in this paper. Although more practically relevant, we assume the code rate $\Rc$ can be adjusted continuously, which is again never the case in practice. Nevertheless, the results in this section can be used to estimate the penalty caused by the use of discrete (and square) constellations.

The selection of code rate and modulation format is assumed to be based on the achievable rates discussed in Sec.~\ref{IT:RatesDCs}. This idea is shown schematically in Fig.~\ref{snr_results}, where horizontal lines with different colors are included. These lines show the SNR ranges where different modulation formats should be used (lines with diamonds for SD-FEC and lines with stars for HD-FEC) and are obtained from Figs.~\ref{OH_vs_SNR_SD} and \ref{OH_vs_SNR_HD}. 

To obtain the throughput achieved by ideal HD- and SD-FEC, we again use \eqref{Th.uniform.final} and follow similar steps to those used in Sec.~\ref{NLIP}. Namely, the SNRs of the routes (shown in Fig.~\ref{snr_results}) are first grouped for each source destination pair and the SNRs are then ``mapped'' to throughputs via the achievable rates discussed in in Sec.~\ref{IT:RatesDCs}. The network throughputs obtained for HD-FEC, are $431$, $217$, and $64$~Tbps, for the DTG, NSF, and GB4 networks, resp. For SD-FEC, these values become $488$, $255$, and $81$~Tbps. These throughput results are shown in the last two columns of Table~\ref{tab:ideal}.

The results in Table~\ref{tab:ideal} show that, when compared to the maximum throughput obtained via the AWGN capacity assumption (fourth column in Table~\ref{tab:ideal}), the use of ideal SD-FEC causes a relative throughput decrease of $9$\%, $8$\%, and $7$\%, for the DTG, NSF, and GB4 networks, resp. This indicates a relatively constant loss across different network topologies. On the other hand, the use of ideal HD-FEC causes relative losses of $18$\%, $22$\%, and $27$\%. These result show an increasing loss as the network size increases, which in turn shows the importance on considering SD-FEC for large networks. We conjecture that these increasing losses are due to the different shape of the ``envelopes'' of the crossing achievable rates in Figs.~\ref{MI_vs_SNR_HD} and \ref{MI_vs_SNR_SD}.

When compared to SD-FEC, HD-FEC codes are typically low complexity and low latency. On the other hand, for the same SNR, HD-FEC codes need higher OH to operate error free, which causes a throughput loss. The relative throughput losses are approximately $12\%$, $15\%$, and $20\%$ for the DTG, NSF, and GB4 networks, resp. This indicates that low complexity and latency can be traded by a $10-20$\% loss in throughput and that the use SD-FEC becomes more and more important as the network size increases.


\section{Practical Schemes}\label{Practical}

Due to the continuous code rate assumption, the throughputs in the last two columns of Table~\ref{tab:ideal} are only upper bounds that cannot be achieved in practice. In this section we discuss practically relevant alternatives.

\subsection{QPSK with $7$\% FEC OH}\label{Sec:QPSKSevenPercent}
	
Probably the simplest (and most popular) alternative in terms of coding and modulation for an optical network is to consider QPSK and a fixed FEC OH of $7$\% across the network. In this case, if the SNR of a given route is below the required SNR for QPSK with $7$\% OH, the route will note be used. If the SNR is above the threshold, then the available throughput in the $r$th route $C_{s,d}^{(r)}$ will be given by the SE in \eqref{SE} times the symbol rate. The total available throughput (in [Tbps]) in \eqref{Th.def} is then given by
\begin{align}
C_{s,d}^{(r)} = 
\begin{cases} 
\frac{0.128}{1.07},	& \text{if $\tnr{SNR}_{s,d}^{(r)}\geq \SNRSevenPercent$}\\
0,													& \text{otherwise}
\end{cases}
\end{align}
where $\tnr{SNR}_{s,d}^{(r)}$ is the SNR of the $r$th route and $\SNRSevenPercent$ is the SNR required for QPSK with $7$\% OH (shown with green pentagons in Figs.~\ref{OH_vs_SNR_HD}, \ref{OH_vs_SNR_SD}, and \ref{snr_results}).

The total network throughput in \eqref{Th.uniform.final} (in Tbps) is then
\begin{align}\label{Th.SevenPercentQPSK}
\Theta = N(N-1) \cdot \frac{0.128}{1.07} \min_{\substack{s,d \in \mathcal{N}\\ d\neq s}} \sum_{r=1}^{R_{s,d}} I_{[\tnr{SNR}_{s,d}^{(r)}\geq \SNRSevenPercent]}
\end{align}
where $I_{[\nu]}$ is an indicator function, i.e., $I_{[\nu]}=1$ if $\nu$ is true, and $I_{[\nu]}=0$ otherwise.

The SNR threshold $\SNRSevenPercent$ in \eqref{Th.SevenPercentQPSK} is different for HD- and SD-FEC, and thus, the resulting throughputs might also be different. However, for both the DTG and NSF networks, all the route SNRs are above both thresholds, and thus, the total network throughput in \eqref{Th.SevenPercentQPSK} is
\begin{align}\label{Th.SevenPercentQPSK.AllGood}
\Theta = N(N-1) \cdot \frac{0.128}{1.07} \min_{\substack{s,d \in \mathcal{N}\\ d\neq s}} \{R_{s,d}\}.
\end{align}
The minimum number of routes for any source destination pair for the NTG and NSF networks are $14$ and $4$, resp., which combined with the number of nodes in the network, give throughputs of $120$ and $87$~Tbps.

For QPSK and 7\% FEC OH, SD offers a theoretical maximum sensitivity increase of about $1.15$~dB\footnote{This can be obtained by comparing the green pentagons for HD and SD FEC in Figs.~\ref{OH_vs_SNR_HD} and \ref{OH_vs_SNR_SD}.}. However, for the DTG and NSF networks, there is no difference between HD and SD in terms of network throughput as the minimum SNR of all routes is above the SNR threshold. This result highlights the fact that under these conditions and traffic assumptions, upgrading all transceivers from HD to SD FEC without changing the OH might not bring any benefit for small networks.

When the GB4 network is considered, however, most of the routes are in fact below the SNR required for QPSK with $7$\% OH (see Fig.~\ref{snr_results}). For this case, the network throughput given by \eqref{Th.SevenPercentQPSK} is in fact zero. This can be intuitively explained by the fact that there are node pairs that are very far apart, and thus, the uniform throughput constraint and full network connectivity cannot be satisfied.

\subsection{One $M$ Schemes}

As an alternative to the QPSK with $7$\% FEC OH approach, in this section we consider two approaches, both of them based on the philosophy that only one modulation format should be implemented across the network.

The first scheme assumes all transceivers implement one modulation format and one code rate. We call this scheme \OROM. In this scheme, the code rate is chosen such that the network throughput (based on achievable rates) is maximized. The results obtained for \OROM~are shown in Fig.~\ref{throughput_results_HD_and_SD_OneM} with red triangles for HD-FEC and with blue triangles for SD-FEC. The top figure shows network throughput, while the bottom ones show optimum code rates. The different networks under consideration are shown from left to right. In the throughput results, we also include the ideal network throughputs in Table~\ref{tab:ideal} (solid horizontal lines) as well as the results obtained with QPSK and 7\% FEC OH from Sec.~\ref{Sec:QPSKSevenPercent} (green pentagons).

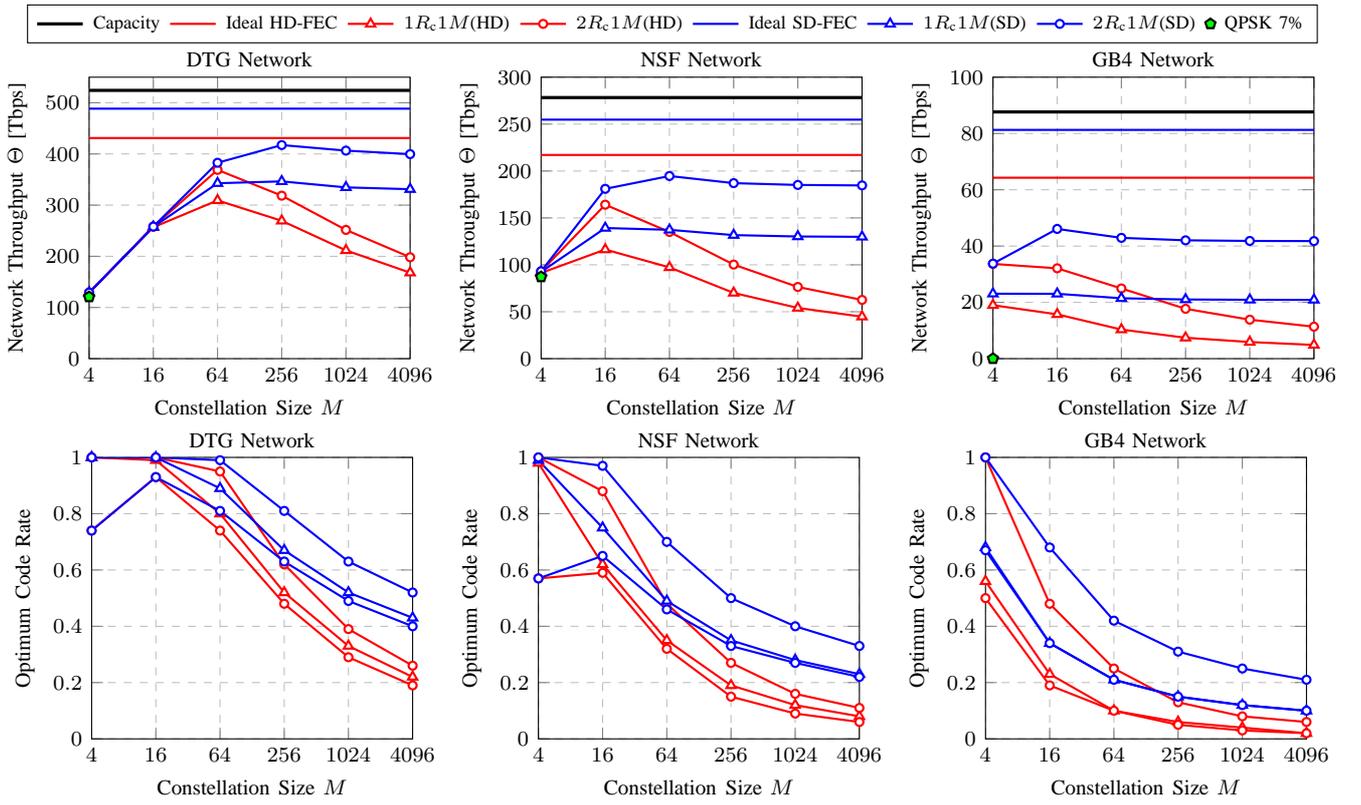
\begin{figure*}
	\begin{center}
		\input{throughput_results_HD_and_SD_OneM.tex}
	\end{center}
	\caption{Throughput (top row) and optimum code rate (bottom row) for one $M$ schemes, for both HD-FEC (red) and SD-FEC (blue), and for the three network topologies in Figs.~\ref{net:DTG}, \ref{net:NSFNET}, and \ref{net:GoogleB4}. The throughput results obtained with QPSK and 7\% FEC OH in Sec.~\ref{Sec:QPSKSevenPercent} are also shown with green pentagons.}
	\label{throughput_results_HD_and_SD_OneM}
\end{figure*}
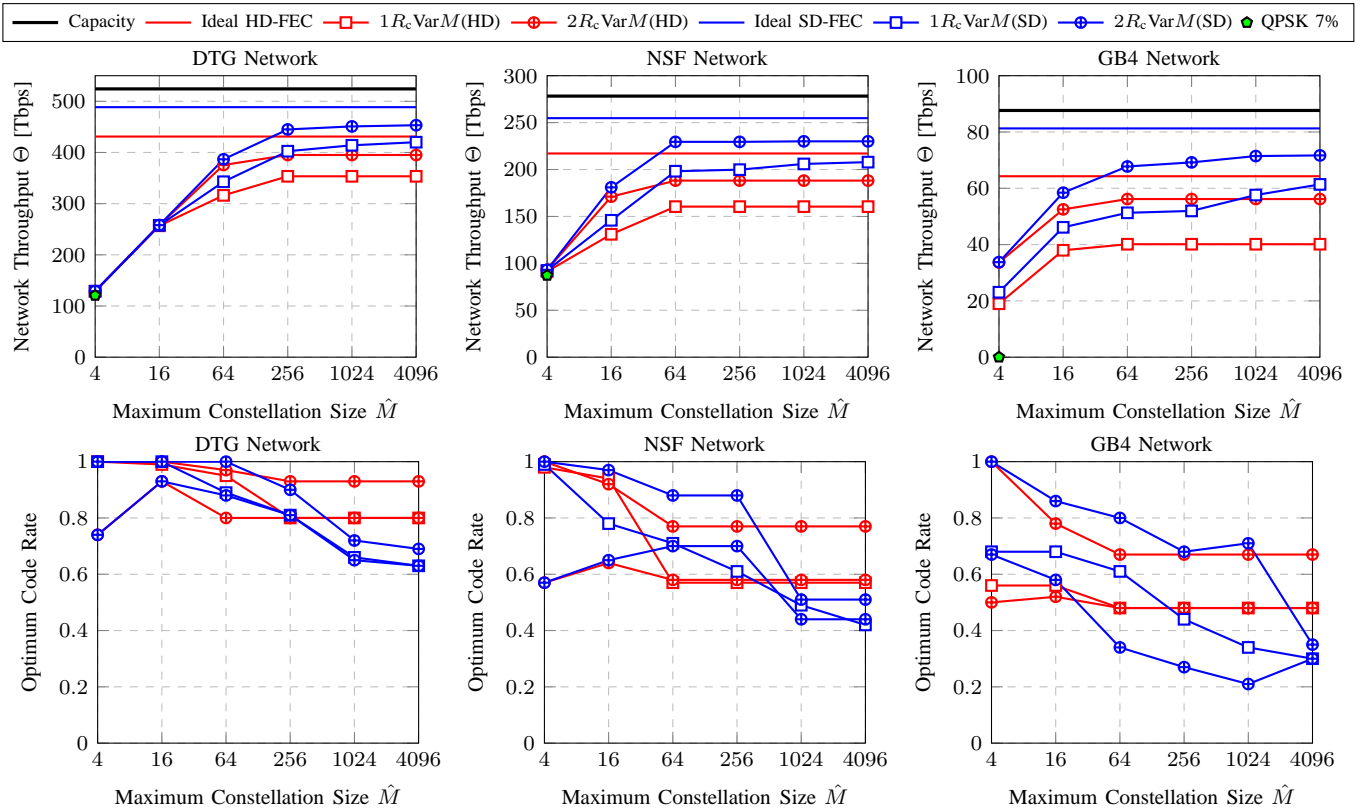
\begin{figure*}
	\begin{center}
		\input{throughput_results_HD_and_SD_VarM.tex}
	\end{center}
	\caption{Throughput (top row) and optimum code rate (bottom row) for variable $M$ schemes, for both HD-FEC (red) and SD-FEC (blue), and for the three network topologies in Figs.~\ref{net:DTG}, \ref{net:NSFNET}, and \ref{net:GoogleB4}. The throughput results obtained with QPSK and 7\% FEC OH in Sec.~\ref{Sec:QPSKSevenPercent} are also shown with green pentagons.}
	\label{throughput_results_HD_and_SD_VarM}
\end{figure*}

The results in Fig.~\ref{throughput_results_HD_and_SD_OneM} show that for HD-FEC and \OROM, there is always a modulation format that is optimum: $M=64$ for DTG, $M=16$ for NSF, and $M=4$ for GB4. Nevertheless, for both HD- and SD-FEC, the gains obtained (with respect to QPSK with $7$\% OH)  by using \OROM~are quite large. Interestingly, the optimum code rates for HD-FEC and \OROM~are $\Rc\approx 0.8$ for DTG, $\Rc\approx 0.62$ for NSF, and $\Rc\approx 0.56$ for GB4 ($24$\%, $61$\% and $79$\% FEC OH, resp.). This indicates a clear potential benefit of using large FEC OH when HD-FEC is used in a network context and where the modulation format is fixed across the network.  

When it comes to \OROM~with SD-FEC, the optimality of a given modulation format is less clear, as the throughput curves in this case do not have a clear peak. This is due to the fact that the crossings between the GMI curves is not as pronounced as the crossings for HD-FEC. Nevertheless, by observing the trend of the curves, a good compromise would be to choose the same modulation formats as for HD-FEC, i.e., $M=64$ for DTG, $M=16$ for NSF, and $M=4$ for GB4. In this case, the corresponding optimum code rates are $\Rc\approx 0.89$, $\Rc\approx 0.75$, and $\Rc\approx 0.68$ ($12$\%, $33$\% and $47$\% FEC OH, resp.). In general, the optimum code rates in this case are slightly higher than the ones for HD-FEC. The gains of SD-FEC over HD-FEC in this case are approximately $50$~Tbps for the DTG and NSF networks. On the other hand, only small gains are observed for the GB4 network.

As mentioned before, the throughput for QPSK with 7\% FEC OH for the GB4 network is zero. On the other hand, for this network, \OROM~gives throughputs of about $20$~Tbps. This is obtained by using QPSK and an increased OH. This result highlights the need for considering high FEC OH in large networks.

The second scheme we consider in this section is called \TROM. In this case, the transceivers are equipped with two code rates but only one modulation format. Again, the code rates are optimized so that the network throughput is maximized.

The results obtained for \TROM~are shown in Fig.~\ref{throughput_results_HD_and_SD_OneM} with circles (red for HD-FEC and blue for SF-FEC). These result show that, regardless of the network under consideration, half of the gap between the ideal FEC limit (horizontal lines) and the throughput obtained by \OROM~can be harvested by using an extra code rate. These results highlight the advantage of adapting the code rate to the variable channel conditions across the network. These results also suggest that a good complexity-performance tradeoff is obtained by using one modulation format and two code rates.

\input{ModulationFormatsHD.tex}

\subsection{Variable $M$ Schemes}

In this section we consider a design alternative where the modulation format is variable. In particular, we assume all the transceivers can choose any modulation format in between QPSK and a given maximum value of $M$, which we denote by $\hat{M}$. For example, if $\hat{M}=64$, then the transceivers can choose from $M=4$, $M=16$ and $M=64$.

The first scheme we consider assumes only one code rate is implemented across the network, and that the code rate is optimized to maximize the network throughput. We denote this scheme \ORVM. The results obtained using this scheme are shown in Fig.~\ref{throughput_results_HD_and_SD_VarM} (squares) and indicate that, in general, constellation sizes beyond $256$QAM give little throughput increases (the throughput curves flatten out for large values of $\hat{M}$). 

Figs.~\ref{ModulationFormatsHD} and \ref{ModulationFormatsSD} show the percentage of transceivers using different modulation formats for \ORVM, for HD- and SD-FEC, resp., and for $\hat{M}=16,64,256$. The value in the middle of each chart is the total throughput obtained. The general trend in these results is that QPSK (green) is only useful in the very large network (GB4). These results also show that, considering $256$QAM (light blue) gives very little throughput increases for the NSF and GB4 networks, for both HD- and SD-FEC. On the other hand, $256$QAM gives a relevant throughput increase for the DTG network.

\input{ModulationFormatsSD.tex}

To compare the throughput contribution of different modulation formats for \ORVM, we show in Fig.~\ref{ORVarMHDandSD} the obtained throughputs for both HD- and SD-FEC. Apart from showing the relative contributions, this figure also shows how the contributions change when SD-FEC is considered instead of HD-FEC. The throughput gains due to SD-FEC are also clearly visible.

\begin{figure}[tbhp]
	\begin{center}
		\input{ORVarMHDandSD.tex}
	\end{center}
	\caption{Network throughput for \ORVM~for HD-FEC (red lines with squares) and SD-FEC (blue lines with squares) as a function of the maximum constellation size for the three network topologies in Figs.~\ref{net:DTG}, \ref{net:NSFNET}, and \ref{net:GoogleB4}. The colors represent the throughput contribution from the different modulation sizes. The lines with markers are the network throughput shown in Fig.~\ref{throughput_results_HD_and_SD_VarM}.}
	\label{ORVarMHDandSD}
\end{figure}

The results in Figs.~\ref{throughput_results_HD_and_SD_VarM}, \ref{ModulationFormatsHD}, \ref{ModulationFormatsSD}, and \ref{ORVarMHDandSD} show that \ORVM~gives similar throughput results to those obtained by \TROM~(two code rates and one modulation format) shown in Fig.~\ref{throughput_results_HD_and_SD_OneM}. This indicates that similar (large) gains, can be obtained by having either multiple code rates or multiple modulation formats.

The second scheme we consider in this section is one where two code rates are implemented across the network (and the modulation can be varied too). We call this scheme \TRVM. The results are shown with plus-circles in Fig.~\ref{throughput_results_HD_and_SD_VarM}. These results indicate that including a second code rate gives a clear advantage with respect to  \ORVM~(squares). This gain is particularly visible for large values of $\hat{M}$ and large networks. In particular, for the GB4 network and $\hat{M}=256$, the gains are approximately $15$~Tbps for both HD- and SD-FEC.

We conclude this section by comparing the optimal code rates of two-rate versus the one-rate schemes. In particular, we observe from Figs.~\ref{throughput_results_HD_and_SD_OneM} and \ref{throughput_results_HD_and_SD_VarM} that the smallest code rate for the two-rate schemes is always quite close to the optimal code rate of the corresponding one-rate scheme. The intuition behind this is that when the transmitters are equipped with two code rates, one low code rate can be used for the worst performing connection, while the other rate is used to increase the overall network throughput.
 
%

\section{Conclusions}\label{Conc}

Optimal constellation sizes and FEC overheads for optical networks were studied. Joint optimization of the constellation and FEC OHs was shown to yield large gains in terms of overall network throughput. The optimal values were shown to be dependent on the SNR distributions within a network. Two code rates and a single constellation (which varies as a function of the network size) gave an good throughput-complexity tradeoff.

In this paper we studied the problem from the point of view of largest achievable rates. Practical FEC implementations, however, will operate a few decibels (or fractions of decibels) away from these achievable rates. Nevertheless, if these penalties are known a priori, the methodology used in this paper can be straightforwardly used to consider practical codes. This is left for future investigation.

\bibliographystyle{IEEEtran}
\bibliography{IEEEabrv,references_all}

\end{document}

%% file: DT2015.tikz
\draw[thin] (105.68pt,56.05pt) -- (-19.29pt,52.71pt) ; 
\node[fill=blue,draw=black,thin,regular polygon, regular polygon sides=3,inner sep=0.4pt,rotate=-208.47] at (80.69pt,55.39pt) {}; 
\node[fill=blue,draw=black,thin,regular polygon, regular polygon sides=3,inner sep=0.4pt,rotate=-208.47] at (55.69pt,54.72pt) {}; 
\node[fill=blue,draw=black,thin,regular polygon, regular polygon sides=3,inner sep=0.4pt,rotate=-208.47] at (30.70pt,54.05pt) {}; 
\node[fill=blue,draw=black,thin,regular polygon, regular polygon sides=3,inner sep=0.4pt,rotate=-208.47] at (5.71pt,53.38pt) {}; 
\draw[thin] (105.68pt,56.05pt) -- (-10.73pt,90.00pt) ; 
\node[fill=blue,draw=black,thin,regular polygon, regular polygon sides=3,inner sep=0.4pt,rotate=133.74] at (82.40pt,62.84pt) {}; 
\node[fill=blue,draw=black,thin,regular polygon, regular polygon sides=3,inner sep=0.4pt,rotate=133.74] at (59.12pt,69.63pt) {}; 
\node[fill=blue,draw=black,thin,regular polygon, regular polygon sides=3,inner sep=0.4pt,rotate=133.74] at (35.84pt,76.42pt) {}; 
\node[fill=blue,draw=black,thin,regular polygon, regular polygon sides=3,inner sep=0.4pt,rotate=133.74] at (12.55pt,83.21pt) {}; 
\draw[thin] (105.68pt,56.05pt) -- (74.01pt,17.59pt) ; 
\node[fill=blue,draw=black,thin,regular polygon, regular polygon sides=3,inner sep=0.4pt,rotate=-159.47] at (95.12pt,43.23pt) {}; 
\node[fill=blue,draw=black,thin,regular polygon, regular polygon sides=3,inner sep=0.4pt,rotate=-159.47] at (84.57pt,30.41pt) {}; 
\draw[thin] (-122.00pt,12.58pt) -- (-56.26pt,-24.21pt) ; 
\node[fill=blue,draw=black,thin,regular polygon, regular polygon sides=3,inner sep=0.4pt,rotate=-59.23] at (-100.09pt,0.32pt) {}; 
\node[fill=blue,draw=black,thin,regular polygon, regular polygon sides=3,inner sep=0.4pt,rotate=-59.23] at (-78.18pt,-11.95pt) {}; 
\draw[thin] (-122.00pt,12.58pt) -- (-19.29pt,52.71pt) ; 
\node[fill=blue,draw=black,thin,regular polygon, regular polygon sides=3,inner sep=0.4pt,rotate=-8.66] at (-101.46pt,20.60pt) {}; 
\node[fill=blue,draw=black,thin,regular polygon, regular polygon sides=3,inner sep=0.4pt,rotate=-8.66] at (-80.91pt,28.63pt) {}; 
\node[fill=blue,draw=black,thin,regular polygon, regular polygon sides=3,inner sep=0.4pt,rotate=-8.66] at (-60.37pt,36.66pt) {}; 
\node[fill=blue,draw=black,thin,regular polygon, regular polygon sides=3,inner sep=0.4pt,rotate=-8.66] at (-39.83pt,44.68pt) {}; 
\draw[thin] (-56.26pt,-24.21pt) -- (-19.29pt,52.71pt) ; 
\node[fill=blue,draw=black,thin,regular polygon, regular polygon sides=3,inner sep=0.4pt,rotate=34.33] at (-48.87pt,-8.83pt) {}; 
\node[fill=blue,draw=black,thin,regular polygon, regular polygon sides=3,inner sep=0.4pt,rotate=34.33] at (-41.47pt,6.56pt) {}; 
\node[fill=blue,draw=black,thin,regular polygon, regular polygon sides=3,inner sep=0.4pt,rotate=34.33] at (-34.08pt,21.94pt) {}; 
\node[fill=blue,draw=black,thin,regular polygon, regular polygon sides=3,inner sep=0.4pt,rotate=34.33] at (-26.68pt,37.33pt) {}; 
\draw[thin] (-56.26pt,-24.21pt) -- (-10.73pt,90.00pt) ; 
\node[fill=blue,draw=black,thin,regular polygon, regular polygon sides=3,inner sep=0.4pt,rotate=38.26] at (-49.76pt,-7.89pt) {}; 
\node[fill=blue,draw=black,thin,regular polygon, regular polygon sides=3,inner sep=0.4pt,rotate=38.26] at (-43.25pt,8.42pt) {}; 
\node[fill=blue,draw=black,thin,regular polygon, regular polygon sides=3,inner sep=0.4pt,rotate=38.26] at (-36.75pt,24.74pt) {}; 
\node[fill=blue,draw=black,thin,regular polygon, regular polygon sides=3,inner sep=0.4pt,rotate=38.26] at (-30.24pt,41.05pt) {}; 
\node[fill=blue,draw=black,thin,regular polygon, regular polygon sides=3,inner sep=0.4pt,rotate=38.26] at (-23.74pt,57.37pt) {}; 
\node[fill=blue,draw=black,thin,regular polygon, regular polygon sides=3,inner sep=0.4pt,rotate=38.26] at (-17.23pt,73.68pt) {}; 
\draw[thin] (-56.26pt,-24.21pt) -- (74.01pt,17.59pt) ; 
\node[fill=blue,draw=black,thin,regular polygon, regular polygon sides=3,inner sep=0.4pt,rotate=-12.21] at (-34.55pt,-17.24pt) {}; 
\node[fill=blue,draw=black,thin,regular polygon, regular polygon sides=3,inner sep=0.4pt,rotate=-12.21] at (-12.84pt,-10.27pt) {}; 
\node[fill=blue,draw=black,thin,regular polygon, regular polygon sides=3,inner sep=0.4pt,rotate=-12.21] at (8.87pt,-3.31pt) {}; 
\node[fill=blue,draw=black,thin,regular polygon, regular polygon sides=3,inner sep=0.4pt,rotate=-12.21] at (30.59pt,3.66pt) {}; 
\node[fill=blue,draw=black,thin,regular polygon, regular polygon sides=3,inner sep=0.4pt,rotate=-12.21] at (52.30pt,10.63pt) {}; 
\draw[thin] (-56.26pt,-24.21pt) -- (42.34pt,-89.42pt) ; 
\node[fill=blue,draw=black,thin,regular polygon, regular polygon sides=3,inner sep=0.4pt,rotate=-63.48] at (-39.83pt,-35.08pt) {}; 
\node[fill=blue,draw=black,thin,regular polygon, regular polygon sides=3,inner sep=0.4pt,rotate=-63.48] at (-23.40pt,-45.95pt) {}; 
\node[fill=blue,draw=black,thin,regular polygon, regular polygon sides=3,inner sep=0.4pt,rotate=-63.48] at (-6.96pt,-56.82pt) {}; 
\node[fill=blue,draw=black,thin,regular polygon, regular polygon sides=3,inner sep=0.4pt,rotate=-63.48] at (9.47pt,-67.69pt) {}; 
\node[fill=blue,draw=black,thin,regular polygon, regular polygon sides=3,inner sep=0.4pt,rotate=-63.48] at (25.91pt,-78.55pt) {}; 
\draw[thin] (-56.26pt,-24.21pt) -- (25.22pt,-45.95pt) ; 
\node[fill=blue,draw=black,thin,regular polygon, regular polygon sides=3,inner sep=0.4pt,rotate=-44.94] at (-29.10pt,-31.46pt) {}; 
\node[fill=blue,draw=black,thin,regular polygon, regular polygon sides=3,inner sep=0.4pt,rotate=-44.94] at (-1.94pt,-38.70pt) {}; 
\draw[thin] (-56.26pt,-24.21pt) -- (-38.97pt,-69.36pt) ; 
\node[fill=blue,draw=black,thin,regular polygon, regular polygon sides=3,inner sep=0.4pt,rotate=-99.05] at (-50.50pt,-39.26pt) {}; 
\node[fill=blue,draw=black,thin,regular polygon, regular polygon sides=3,inner sep=0.4pt,rotate=-99.05] at (-44.74pt,-54.31pt) {}; 
\draw[thin] (-19.29pt,52.71pt) -- (-10.73pt,90.00pt) ; 
\node[fill=blue,draw=black,thin,regular polygon, regular polygon sides=3,inner sep=0.4pt,rotate=47.07] at (-15.01pt,71.36pt) {}; 
\draw[thin] (-19.29pt,52.71pt) -- (74.01pt,17.59pt) ; 
\node[fill=blue,draw=black,thin,regular polygon, regular polygon sides=3,inner sep=0.4pt,rotate=-50.63] at (4.04pt,43.93pt) {}; 
\node[fill=blue,draw=black,thin,regular polygon, regular polygon sides=3,inner sep=0.4pt,rotate=-50.63] at (27.36pt,35.15pt) {}; 
\node[fill=blue,draw=black,thin,regular polygon, regular polygon sides=3,inner sep=0.4pt,rotate=-50.63] at (50.69pt,26.37pt) {}; 
\draw[thin] (74.01pt,17.59pt) -- (25.22pt,-45.95pt) ; 
\node[fill=blue,draw=black,thin,regular polygon, regular polygon sides=3,inner sep=0.4pt,rotate=-157.52] at (61.81pt,1.71pt) {}; 
\node[fill=blue,draw=black,thin,regular polygon, regular polygon sides=3,inner sep=0.4pt,rotate=-157.52] at (49.62pt,-14.18pt) {}; 
\node[fill=blue,draw=black,thin,regular polygon, regular polygon sides=3,inner sep=0.4pt,rotate=-157.52] at (37.42pt,-30.06pt) {}; 
\draw[thin] (42.34pt,-89.42pt) -- (25.22pt,-45.95pt) ; 
\node[fill=blue,draw=black,thin,regular polygon, regular polygon sides=3,inner sep=0.4pt,rotate=81.49] at (36.63pt,-74.93pt) {}; 
\node[fill=blue,draw=black,thin,regular polygon, regular polygon sides=3,inner sep=0.4pt,rotate=81.49] at (30.93pt,-60.44pt) {}; 
\draw[thin] (42.34pt,-89.42pt) -- (-38.97pt,-69.36pt) ; 
\node[fill=blue,draw=black,thin,regular polygon, regular polygon sides=3,inner sep=0.4pt,rotate=136.14] at (22.01pt,-84.41pt) {}; 
\node[fill=blue,draw=black,thin,regular polygon, regular polygon sides=3,inner sep=0.4pt,rotate=136.14] at (1.68pt,-79.39pt) {}; 
\node[fill=blue,draw=black,thin,regular polygon, regular polygon sides=3,inner sep=0.4pt,rotate=136.14] at (-18.65pt,-74.37pt) {}; 
\draw[thin] (25.22pt,-45.95pt) -- (-38.97pt,-69.36pt) ; 
\node[fill=blue,draw=black,thin,regular polygon, regular polygon sides=3,inner sep=0.4pt,rotate=-189.96] at (3.82pt,-53.75pt) {}; 
\node[fill=blue,draw=black,thin,regular polygon, regular polygon sides=3,inner sep=0.4pt,rotate=-189.96] at (-17.58pt,-61.55pt) {}; 
\filldraw[fill=figgreen,draw=black,thin] (105.68pt,56.05pt) circle (3.5pt) node[black] {\tiny $1$}; 
\filldraw[fill=figgreen,draw=black,thin] (-122.00pt,12.58pt) circle (3.5pt) node[black] {\tiny $2$}; 
\filldraw[fill=figgreen,draw=black,thin] (-56.26pt,-24.21pt) circle (3.5pt) node[black] {\tiny $3$}; 
\filldraw[fill=figgreen,draw=black,thin] (-19.29pt,52.71pt) circle (3.5pt) node[black] {\tiny $4$}; 
\filldraw[fill=figgreen,draw=black,thin] (-10.73pt,90.00pt) circle (3.5pt) node[black] {\tiny $5$}; 
\filldraw[fill=figgreen,draw=black,thin] (74.01pt,17.59pt) circle (3.5pt) node[black] {\tiny $6$}; 
\filldraw[fill=figgreen,draw=black,thin] (42.34pt,-89.42pt) circle (3.5pt) node[black] {\tiny $7$}; 
\filldraw[fill=figgreen,draw=black,thin] (25.22pt,-45.95pt) circle (3.5pt) node[black] {\tiny $8$}; 
\filldraw[fill=figgreen,draw=black,thin] (-38.97pt,-69.36pt) circle (3.5pt) node[black] {\tiny $9$}; 

%% file: GoogleB4.tikz
\draw[thin] (179.60pt,71.17pt) -- (142.32pt,38.50pt) ; 
\draw[thin] (179.60pt,71.17pt) -- (139.98pt,33.37pt) ; 
\draw[thin] (142.32pt,38.50pt) -- (139.98pt,33.37pt) ; 
\draw[thin] (142.32pt,38.50pt) -- (126.25pt,47.86pt) ; 
\draw[thin] (142.32pt,38.50pt) -- (8.61pt,-17.22pt) ; 
\draw[thin] (139.98pt,33.37pt) -- (126.25pt,47.86pt) ; 
\draw[thin] (139.98pt,33.37pt) -- (6.21pt,-8.40pt) ; 
\draw[thin] (8.61pt,-17.22pt) -- (6.21pt,-8.40pt) ; 
\draw[thin] (8.61pt,-17.22pt) -- (4.12pt,-22.64pt) ; 
\draw[thin] (8.61pt,-17.22pt) -- (-16.63pt,8.83pt) ; 
\draw[thin] (6.21pt,-8.40pt) -- (4.12pt,-22.64pt) ; 
\draw[thin] (6.21pt,-8.40pt) -- (-16.63pt,8.83pt) ; 
\draw[thin] (4.12pt,-22.64pt) -- (-15.75pt,-7.40pt) ; 
\draw[thin] (4.12pt,-22.64pt) -- (-57.00pt,22.84pt) ; 
\draw[thin] (-16.63pt,8.83pt) -- (-15.75pt,-7.40pt) ; 
\draw[thin] (-16.63pt,8.83pt) -- (-57.00pt,22.84pt) ; 
\draw[thin] (-15.75pt,-7.40pt) -- (-245.71pt,-46.92pt) ; 
\draw[thin] (-57.00pt,22.84pt) -- (-272.00pt,-120.00pt) ; 
\draw[thin] (-245.71pt,-46.92pt) -- (-272.00pt,-120.00pt) ; 
\filldraw[fill=figgreen,draw=black,thin] (179.60pt,71.17pt) circle (3.5pt) node[black] {\tiny $1$}; 
\filldraw[fill=figgreen,draw=black,thin] (142.32pt,38.50pt) circle (3.5pt) node[black] {\tiny $2$}; 
\filldraw[fill=figgreen,draw=black,thin] (139.98pt,33.37pt) circle (3.5pt) node[black] {\tiny $3$}; 
\filldraw[fill=figgreen,draw=black,thin] (126.25pt,47.86pt) circle (3.5pt) node[black] {\tiny $4$}; 
\filldraw[fill=figgreen,draw=black,thin] (8.61pt,-17.22pt) circle (3.5pt) node[black] {\tiny $5$}; 
\filldraw[fill=figgreen,draw=black,thin] (6.21pt,-8.40pt) circle (3.5pt) node[black] {\tiny $6$}; 
\filldraw[fill=figgreen,draw=black,thin] (4.12pt,-22.64pt) circle (3.5pt) node[black] {\tiny $7$}; 
\filldraw[fill=figgreen,draw=black,thin] (-16.63pt,8.83pt) circle (3.5pt) node[black] {\tiny $8$}; 
\filldraw[fill=figgreen,draw=black,thin] (-15.75pt,-7.40pt) circle (3.5pt) node[black] {\tiny $9$}; 
\filldraw[fill=figgreen,draw=black,thin] (-57.00pt,22.84pt) circle (3.5pt) node[black] {\tiny $10$}; 
\filldraw[fill=figgreen,draw=black,thin] (-245.71pt,-46.92pt) circle (3.5pt) node[black] {\tiny $11$}; 
\filldraw[fill=figgreen,draw=black,thin] (-272.00pt,-120.00pt) circle (3.5pt) node[black] {\tiny $12$}; 

%% file: snr_results.tex
\pgfplotscreateplotcyclelist{color list}{brown,green,blue,red,cyan,purple,green,blue,red,cyan,purple}
\begin{tikzpicture}
\begin{axis}[cycle list name=color list,
	xminorgrids=true,
        width=.35\textwidth,
        height=0.6\columnwidth,
        grid=both,
        xmin=0,xmax=27,
        ymin=0,ymax=220,
        xlabel={$\tnr{SNR}$~[dB]},
        xlabel style={yshift=0.1cm},
        ylabel={Number of Transceivers},
        ylabel style={yshift=-0.1cm},
        xtick={0,5,...,30},
        every axis/.append style={font=\footnotesize},
	grid style={dashed},
	title=DTG Network,
	title style={yshift=-0.2cm}]
\addplot+[ycomb,very thin,mark size=1pt,mark=triangle*] file {SNR_dist_DT2015.dat};
\snrhorizontalbarsHD{190}
\snrhorizontalbarsSD{210}
\snrcrossingsHD{210}{190}
\addplot[color=black,mark=pentagon*,fill=green,thick,only marks,mark size=2pt] coordinates {(7.68,190)};
\addplot[color=black,mark=pentagon*,fill=green,thick,only marks,mark size=2pt] coordinates {(6.56,210)};
\end{axis}
\end{tikzpicture}
\begin{tikzpicture}
\begin{axis}[cycle list name=color list,
	xminorgrids=true,
        width=.35\textwidth,
        height=0.6\columnwidth,
        grid=both,
        xmin=0,xmax=27,
        ymin=0,ymax=50,
        xlabel={$\tnr{SNR}$~[dB]},
        xlabel style={yshift=0.1cm},
        ylabel={Number of Transceivers},
        ylabel style={yshift=-0.1cm},
        xtick={0,5,...,30},
        every axis/.append style={font=\footnotesize},
	grid style={dashed},
	title=NSF Network,
	title style={yshift=-0.2cm}]
\addplot+[ycomb,very thin,mark size=1pt,mark=triangle*] file {SNR_dist_NSFNET.dat};
\snrhorizontalbarsHD{44}
\snrhorizontalbarsSD{48}
\snrcrossingsHD{48}{44}
\addplot[color=black,mark=pentagon*,fill=green,thick,only marks,mark size=2pt] coordinates {(7.68,44)};
\addplot[color=black,mark=pentagon*,fill=green,thick,only marks,mark size=2pt] coordinates {(6.56,48)};
\end{axis}
\end{tikzpicture}
\begin{tikzpicture}
\begin{axis}[cycle list name=color list,
	xminorgrids=true,
        width=.35\textwidth,
        height=0.6\columnwidth,
        grid=both,
        xmin=0,xmax=27,
        ymin=0,ymax=30,
        xlabel={$\tnr{SNR}$~[dB]},
        xlabel style={yshift=0.1cm},
        ylabel={Number of Transceivers},
        ylabel style={yshift=-0.1cm},
        xtick={0,5,...,30},
        every axis/.append style={font=\footnotesize},
	grid style={dashed},
	title=GB4 Network,
	title style={yshift=-0.2cm}]
\addplot+[ycomb,very thin,mark size=1pt,mark=triangle*] file {SNR_dist_GoogleB4.dat};
\snrhorizontalbarsHD{26}
\snrhorizontalbarsSD{28.8}
\snrcrossingsHD{28.8}{26}
\addplot[color=black,mark=pentagon*,fill=green,thick,only marks,mark size=2pt] coordinates {(7.68,26)};
\addplot[color=black,mark=pentagon*,fill=green,thick,only marks,mark size=2pt] coordinates {(6.56,28.8)};
\end{axis}
\end{tikzpicture}

%% file: throughput_results_HD_and_SD_OneM.tex
\pgfplotsset{compat=1.3}
\ref{legendnameTOneM}
\begin{tikzpicture}
\begin{axis}[
		legend columns=-1,
        legend entries={Capacity, Ideal HD-FEC, \OROM (HD), \TROM (HD), Ideal SD-FEC, \OROM (SD), \TROM (SD), QPSK 7\%},
		legend style={font=\scriptsize,column sep=0.05cm},
        legend to name=legendnameTOneM,
		xminorgrids=true,
        width=.32\textwidth,
        height=0.6\columnwidth,
        grid=both,
        xmin=1,xmax=6,
        xtick={1,2,...,6},
        xticklabels={$4$,$16$,$64$,$256$,$1024$,$4096$},
        ymin=0,ymax=550,
        ytick={0,100,...,550},
        xlabel={Constellation Size $M$},
        ylabel={Network Throughput $\Theta$~[Tbps]},
        every axis/.append style={font=\footnotesize},
		grid style={dashed},
		title=DTG Network,
		title style={yshift=-0.2cm}]
\addplot[color=black,very thick] file {Throughputs_Capacity_DT2015.dat};
\addplot[color=red, thick] file {Throughputs_CRCMHD_DT2015.dat};
\addplot[color=red,mark=triangle*, thick,mark options={fill=white}] file {Throughputs_OROMHD_DT2015.dat};
\addplot[color=red,mark=*, thick,mark options={fill=white},mark size=1.5pt] file {Throughputs_TROMHD_DT2015.dat};
\addplot[color=blue, thick] file {Throughputs_CRCMSD_DT2015.dat};
\addplot[color=blue,mark=triangle*, thick,mark options={fill=white}] file {Throughputs_OROMSD_DT2015.dat};
\addplot[color=blue,mark=*, thick,mark options={fill=white},mark size=1.5pt] file {Throughputs_TROMSD_DT2015.dat};
\addplot[color=black,mark=pentagon*,fill=green,thick,only marks,mark size=2pt] file {Throughputs_CRCMSDSevenPercentQPSK_DT2015.dat};
\end{axis}
\end{tikzpicture}
\begin{tikzpicture}
\begin{axis}[
	xminorgrids=true,
        width=.32\textwidth,
        height=0.6\columnwidth,
        grid=both,
        xmin=1,xmax=6,
        xtick={1,2,...,6},
        xticklabels={$4$,$16$,$64$,$256$,$1024$,$4096$},
        ymin=0,ymax=300,
        ytick={0,50,...,300},
        xlabel={Constellation Size $M$},
        ylabel={Network Throughput $\Theta$~[Tbps]},
        every axis/.append style={font=\footnotesize},
		grid style={dashed},
		title=NSF Network,
		title style={yshift=-0.2cm}]
\addplot[color=black,very thick] file {Throughputs_Capacity_NSFNET.dat};
\addplot[color=red, thick] file {Throughputs_CRCMHD_NSFNET.dat};
\addplot[color=red,mark=triangle*, thick,mark options={fill=white}] file {Throughputs_OROMHD_NSFNET.dat};
\addplot[color=red,mark=*, thick,mark options={fill=white},mark size=1.5pt] file {Throughputs_TROMHD_NSFNET.dat};
\addplot[color=blue, thick] file {Throughputs_CRCMSD_NSFNET.dat};
\addplot[color=blue,mark=triangle*, thick,mark options={fill=white}] file {Throughputs_OROMSD_NSFNET.dat};
\addplot[color=blue,mark=*, thick,mark options={fill=white},mark size=1.5pt] file {Throughputs_TROMSD_NSFNET.dat};
\addplot[color=black,mark=pentagon*,fill=green,thick,only marks,mark size=2pt] file {Throughputs_CRCMSDSevenPercentQPSK_NSFNET.dat};
\end{axis}
\end{tikzpicture}
\begin{tikzpicture}
\begin{axis}[
	xminorgrids=true,
        width=.32\textwidth,
        height=0.6\columnwidth,
        grid=both,
        xmin=1,xmax=6,
        xtick={1,2,...,6},
        xticklabels={$4$,$16$,$64$,$256$,$1024$,$4096$},
        ymin=0,ymax=100,
        ytick={0,20,...,100},
        xlabel={Constellation Size $M$},
        ylabel={Network Throughput $\Theta$~[Tbps]},
        every axis/.append style={font=\footnotesize},
		grid style={dashed},
		title=GB4 Network,
		title style={yshift=-0.2cm}]
\addplot[color=black,very thick] file {Throughputs_Capacity_GoogleB4.dat};
\addplot[color=red, thick] file {Throughputs_CRCMHD_GoogleB4.dat};
\addplot[color=red,mark=triangle*,mark options={fill=white}, thick] file {Throughputs_OROMHD_GoogleB4.dat};
\addplot[color=red,mark=*, thick,mark options={fill=white},mark size=1.5pt] file {Throughputs_TROMHD_GoogleB4.dat};
\addplot[color=blue, thick] file {Throughputs_CRCMSD_GoogleB4.dat};
\addplot[color=blue,mark=triangle*,mark options={fill=white}, thick] file {Throughputs_OROMSD_GoogleB4.dat};
\addplot[color=blue,mark=*, thick,mark options={fill=white},mark size=1.5pt] file {Throughputs_TROMSD_GoogleB4.dat};
\addplot[color=black,mark=pentagon*,fill=green,thick,only marks,mark size=2pt] file {Throughputs_CRCMHDSevenPercentQPSK_GoogleB4.dat};
\end{axis}
\end{tikzpicture}
\begin{tikzpicture}
\begin{axis}[
	xminorgrids=true,
        width=.32\textwidth,
        height=0.6\columnwidth,
        grid=both,
        xmin=1,xmax=6,
        xtick={1,2,...,6},
        xticklabels={$4$,$16$,$64$,$256$,$1024$,$4096$},
        ymin=0,ymax=1,
        ytick={0,0.2,...,1},
        xlabel={Constellation Size $M$},
        ylabel={Optimum Code Rate},
        every axis/.append style={font=\footnotesize},
	legend entries={},
	legend style={legend pos=north east,font=\tiny,legend cell align=left},
	grid style={dashed},
	title=DTG Network,
	title style={yshift=-0.2cm}]
\addplot[color=red,mark=triangle*, thick,mark options={fill=white}] file {OptimumR_OROMHD_DT2015.dat};
\addplot[color=red,mark=*, thick,mark options={fill=white},mark size=1.5pt] file {OptimumR1_TROMHD_DT2015.dat};
\addplot[color=red,mark=*, thick,mark options={fill=white},mark size=1.5pt] file {OptimumR2_TROMHD_DT2015.dat};
\addplot[color=blue,mark=triangle*, thick,mark options={fill=white}] file {OptimumR_OROMSD_DT2015.dat};
\addplot[color=blue,mark=*, thick,mark options={fill=white},mark size=1.5pt] file {OptimumR1_TROMSD_DT2015.dat};
\addplot[color=blue,mark=*, thick,mark options={fill=white},mark size=1.5pt] file {OptimumR2_TROMSD_DT2015.dat};
\end{axis}
\end{tikzpicture}
\begin{tikzpicture}
\begin{axis}[
	xminorgrids=true,
        width=.32\textwidth,
        height=0.6\columnwidth,
        grid=both,
        xmin=1,xmax=6,
        xtick={1,2,...,6},
        xticklabels={$4$,$16$,$64$,$256$,$1024$,$4096$},
        ymin=0,ymax=1,
        ytick={0,0.2,...,1},
        xlabel={Constellation Size $M$},
        ylabel={Optimum Code Rate},
        every axis/.append style={font=\footnotesize},
	grid style={dashed},
	title=NSF Network,
	title style={yshift=-0.2cm}]
\addplot[color=red,mark=triangle*,mark options={fill=white}, thick] file {OptimumR_OROMHD_NSFNET.dat};
\addplot[color=red,mark=*, thick,mark options={fill=white},mark size=1.5pt] file {OptimumR1_TROMHD_NSFNET.dat};
\addplot[color=red,mark=*, thick,mark options={fill=white},mark size=1.5pt] file {OptimumR2_TROMHD_NSFNET.dat};
\addplot[color=blue,mark=triangle*,mark options={fill=white}, thick] file {OptimumR_OROMSD_NSFNET.dat};
\addplot[color=blue,mark=*, thick,mark options={fill=white},mark size=1.5pt] file {OptimumR1_TROMSD_NSFNET.dat};
\addplot[color=blue,mark=*, thick,mark options={fill=white},mark size=1.5pt] file {OptimumR2_TROMSD_NSFNET.dat};
\end{axis}
\end{tikzpicture}
\begin{tikzpicture}
\begin{axis}[
	xminorgrids=true,
        width=.32\textwidth,
        height=0.6\columnwidth,
        grid=both,
        xmin=1,xmax=6,
        xtick={1,2,...,6},
        xticklabels={$4$,$16$,$64$,$256$,$1024$,$4096$},
        ymin=0,ymax=1,
        ytick={0,0.2,...,1},
        xlabel={Constellation Size $M$},
        ylabel={Optimum Code Rate},
        every axis/.append style={font=\footnotesize},
	grid style={dashed},
	title=GB4 Network,
	title style={yshift=-0.2cm}]
\addplot[color=red,mark=triangle*,mark options={fill=white}, thick] file {OptimumR_OROMHD_GoogleB4.dat};
\addplot[color=red,mark=*, thick,mark options={fill=white},mark size=1.5pt] file {OptimumR1_TROMHD_GoogleB4.dat};
\addplot[color=red,mark=*, thick,mark options={fill=white},mark size=1.5pt] file {OptimumR2_TROMHD_GoogleB4.dat};
\addplot[color=blue,mark=triangle*,mark options={fill=white}, thick] file {OptimumR_OROMSD_GoogleB4.dat};
\addplot[color=blue,mark=*, thick,mark options={fill=white},mark size=1.5pt] file {OptimumR1_TROMSD_GoogleB4.dat};
\addplot[color=blue,mark=*, thick,mark options={fill=white},mark size=1.5pt] file {OptimumR2_TROMSD_GoogleB4.dat};
\end{axis}
\end{tikzpicture}

%% file: throughput_results_HD_and_SD_VarM.tex
\pgfplotsset{compat=1.3}
\ref{legendnameTVarM}
\begin{tikzpicture}
\begin{axis}[
		legend columns=-1,
        legend entries={Capacity, Ideal HD-FEC, \ORVM (HD), \TRVM (HD), Ideal SD-FEC, \ORVM (SD), \TRVM (SD), QPSK 7\%},
		legend style={font=\scriptsize,column sep=0.05cm},
        legend to name=legendnameTVarM,
		xminorgrids=true,
        width=.32\textwidth,
        height=0.6\columnwidth,
        grid=both,
        xmin=1,xmax=6,
        xtick={1,2,...,6},
        xticklabels={$4$,$16$,$64$,$256$,$1024$,$4096$},
        ymin=0,ymax=550,
        ytick={0,100,...,550},
        xlabel={Maximum Constellation Size $\hat{M}$},
        ylabel={Network Throughput $\Theta$~[Tbps]},
        every axis/.append style={font=\footnotesize},
		grid style={dashed},
		title=DTG Network,
		title style={yshift=-0.2cm}]
\addplot[color=black,very thick] file {Throughputs_Capacity_DT2015.dat};
\addplot[color=red, thick] file {Throughputs_CRCMHD_DT2015.dat};
\addplot[color=red,mark=square*, thick,mark options={fill=white}] file {Throughputs_ORVMHD_DT2015.dat};
\addplot[color=red,mark=oplus*, thick,mark options={fill=white},mark size=2pt] file {Throughputs_TRVMHD_DT2015.dat};
\addplot[color=blue, thick] file {Throughputs_CRCMSD_DT2015.dat};
\addplot[color=blue,mark=square*, thick,mark options={fill=white}] file {Throughputs_ORVMSD_DT2015.dat};
\addplot[color=blue,mark=oplus*, thick,mark options={fill=white},mark size=2pt] file {Throughputs_TRVMSD_DT2015.dat};
\addplot[color=black,mark=pentagon*,fill=green,thick,only marks,mark size=2pt] file {Throughputs_CRCMSDSevenPercentQPSK_DT2015.dat};
\end{axis}
\end{tikzpicture}
\begin{tikzpicture}
\begin{axis}[
	xminorgrids=true,
        width=.32\textwidth,
        height=0.6\columnwidth,
        grid=both,
        xmin=1,xmax=6,
        xtick={1,2,...,6},
        xticklabels={$4$,$16$,$64$,$256$,$1024$,$4096$},
        ymin=0,ymax=300,
        ytick={0,50,...,300},
        xlabel={Maximum Constellation Size $\hat{M}$},
        ylabel={Network Throughput $\Theta$~[Tbps]},
        every axis/.append style={font=\footnotesize},
		grid style={dashed},
		title=NSF Network,
		title style={yshift=-0.2cm}]
\addplot[color=black,very thick] file {Throughputs_Capacity_NSFNET.dat};
\addplot[color=red, thick] file {Throughputs_CRCMHD_NSFNET.dat};
\addplot[color=red,mark=square*, thick,mark options={fill=white}] file {Throughputs_ORVMHD_NSFNET.dat};
\addplot[color=red,mark=oplus*, thick,mark options={fill=white},mark size=2pt] file {Throughputs_TRVMHD_NSFNET.dat};
\addplot[color=blue, thick] file {Throughputs_CRCMSD_NSFNET.dat};
\addplot[color=blue,mark=square*, thick,mark options={fill=white}] file {Throughputs_ORVMSD_NSFNET.dat};
\addplot[color=blue,mark=oplus*, thick,mark options={fill=white},mark size=2pt] file {Throughputs_TRVMSD_NSFNET.dat};
\addplot[color=black,mark=pentagon*,fill=green,thick,only marks,mark size=2pt] file {Throughputs_CRCMSDSevenPercentQPSK_NSFNET.dat};
\end{axis}
\end{tikzpicture}
\begin{tikzpicture}
\begin{axis}[
	xminorgrids=true,
        width=.32\textwidth,
        height=0.6\columnwidth,
        grid=both,
        xmin=1,xmax=6,
        xtick={1,2,...,6},
        xticklabels={$4$,$16$,$64$,$256$,$1024$,$4096$},
        ymin=0,ymax=100,
        ytick={0,20,...,100},
        xlabel={Maximum Constellation Size $\hat{M}$},
        ylabel={Network Throughput $\Theta$~[Tbps]},
        every axis/.append style={font=\footnotesize},
		grid style={dashed},
		title=GB4 Network,
		title style={yshift=-0.2cm}]
\addplot[color=black,very thick] file {Throughputs_Capacity_GoogleB4.dat};
\addplot[color=red, thick] file {Throughputs_CRCMHD_GoogleB4.dat};
\addplot[color=red,mark=square*,mark options={fill=white}, thick] file {Throughputs_ORVMHD_GoogleB4.dat};
\addplot[color=red,mark=oplus*, thick,mark options={fill=white},mark size=2pt] file {Throughputs_TRVMHD_GoogleB4.dat};
\addplot[color=blue, thick] file {Throughputs_CRCMSD_GoogleB4.dat};
\addplot[color=blue,mark=square*,mark options={fill=white}, thick] file {Throughputs_ORVMSD_GoogleB4.dat};
\addplot[color=blue,mark=oplus*, thick,mark options={fill=white},mark size=2pt] file {Throughputs_TRVMSD_GoogleB4.dat};
\addplot[color=black,mark=pentagon*,fill=green,thick,only marks,mark size=2pt] file {Throughputs_CRCMHDSevenPercentQPSK_GoogleB4.dat};
\end{axis}
\end{tikzpicture}
\begin{tikzpicture}
\begin{axis}[
	xminorgrids=true,
        width=.32\textwidth,
        height=0.6\columnwidth,
        grid=both,
        xmin=1,xmax=6,
        xtick={1,2,...,6},
        xticklabels={$4$,$16$,$64$,$256$,$1024$,$4096$},
        ymin=0,ymax=1,
        ytick={0,0.2,...,1},
        xlabel={Maximum Constellation Size $\hat{M}$},
        ylabel={Optimum Code Rate},
        every axis/.append style={font=\footnotesize},
	legend entries={},
	legend style={legend pos=north east,font=\tiny,legend cell align=left},
	grid style={dashed},
	title=DTG Network,
	title style={yshift=-0.2cm}]
\addplot[color=red,mark=square*, thick,mark options={fill=white}] file {OptimumR_ORVMHD_DT2015.dat};
\addplot[color=red,mark=oplus*, thick,mark options={fill=white},mark size=2pt] file {OptimumR1_TRVMHD_DT2015.dat};
\addplot[color=red,mark=oplus*, thick,mark options={fill=white},mark size=2pt] file {OptimumR2_TRVMHD_DT2015.dat};
\addplot[color=blue,mark=square*, thick,mark options={fill=white}] file {OptimumR_ORVMSD_DT2015.dat};
\addplot[color=blue,mark=oplus*, thick,mark options={fill=white},mark size=2pt] file {OptimumR1_TRVMSD_DT2015.dat};
\addplot[color=blue,mark=oplus*, thick,mark options={fill=white},mark size=2pt] file {OptimumR2_TRVMSD_DT2015.dat};
\end{axis}
\end{tikzpicture}
\begin{tikzpicture}
\begin{axis}[
	xminorgrids=true,
        width=.32\textwidth,
        height=0.6\columnwidth,
        grid=both,
        xmin=1,xmax=6,
        xtick={1,2,...,6},
        xticklabels={$4$,$16$,$64$,$256$,$1024$,$4096$},
        ymin=0,ymax=1,
        ytick={0,0.2,...,1},
        xlabel={Maximum Constellation Size $\hat{M}$},
        ylabel={Optimum Code Rate},
        every axis/.append style={font=\footnotesize},
	grid style={dashed},
	title=NSF Network,
	title style={yshift=-0.2cm}]
\addplot[color=red,mark=square*,mark options={fill=white}, thick] file {OptimumR_ORVMHD_NSFNET.dat};
\addplot[color=red,mark=oplus*, thick,mark options={fill=white},mark size=2pt] file {OptimumR1_TRVMHD_NSFNET.dat};
\addplot[color=red,mark=oplus*, thick,mark options={fill=white},mark size=2pt] file {OptimumR2_TRVMHD_NSFNET.dat};
\addplot[color=blue,mark=square*,mark options={fill=white}, thick] file {OptimumR_ORVMSD_NSFNET.dat};
\addplot[color=blue,mark=oplus*, thick,mark options={fill=white},mark size=2pt] file {OptimumR1_TRVMSD_NSFNET.dat};
\addplot[color=blue,mark=oplus*, thick,mark options={fill=white},mark size=2pt] file {OptimumR2_TRVMSD_NSFNET.dat};
\end{axis}
\end{tikzpicture}
\begin{tikzpicture}
\begin{axis}[
	xminorgrids=true,
        width=.32\textwidth,
        height=0.6\columnwidth,
        grid=both,
        xmin=1,xmax=6,
        xtick={1,2,...,6},
        xticklabels={$4$,$16$,$64$,$256$,$1024$,$4096$},
        ymin=0,ymax=1,
        ytick={0,0.2,...,1},
        xlabel={Maximum Constellation Size $\hat{M}$},
        ylabel={Optimum Code Rate},
        every axis/.append style={font=\footnotesize},
	grid style={dashed},
	title=GB4 Network,
	title style={yshift=-0.2cm}]
\addplot[color=red,mark=square*,mark options={fill=white}, thick] file {OptimumR_ORVMHD_GoogleB4.dat};
\addplot[color=red,mark=oplus*, thick,mark options={fill=white},mark size=2pt] file {OptimumR1_TRVMHD_GoogleB4.dat};
\addplot[color=red,mark=oplus*, thick,mark options={fill=white},mark size=2pt] file {OptimumR2_TRVMHD_GoogleB4.dat};
\addplot[color=blue,mark=square*,mark options={fill=white}, thick] file {OptimumR_ORVMSD_GoogleB4.dat};
\addplot[color=blue,mark=oplus*, thick,mark options={fill=white},mark size=2pt] file {OptimumR1_TRVMSD_GoogleB4.dat};
\addplot[color=blue,mark=oplus*, thick,mark options={fill=white},mark size=2pt] file {OptimumR2_TRVMSD_GoogleB4.dat};
\end{axis}
\end{tikzpicture}

%% file: ModulationFormatsHD.tex
\begin{figure*}
\begin{center}
\setlength{\tabcolsep}{3pt}
\renewcommand{\arraystretch}{1.4}
{\footnotesize
\begin{tabular}{cccc}
\hline 

\hline
Mod. Size & $\hat{M}=16$ & $\hat{M}=64$ & $\hat{M}=256$\\
\hline 

\hline
\rotatebox{90}{\hspace{.65cm} DGT Network}
&\wheelchart{422/green/QSPK, 606/blue/$16$QAM}{20}{257}
&\wheelchart{316/blue/$16$QAM, 299/red/$64$QAM}{20}{316}
&\wheelchart{9/blue/$16$QAM, 503/red/$64$QAM, 103/cyan/$256$QAM}{75}{353}\\
\hline 
\rotatebox{90}{\hspace{.65cm} NSF Network}
&\wheelchart{422/green/QSPK, 125/blue/$16$QAM}{60}{131}
&\wheelchart{1/green/QSPK, 380/blue/$16$QAM, 166/red/$64$QAM}{75}{160}
&\wheelchart{1/green/QSPK, 380/blue/$16$QAM, 150/red/$64$QAM, 16/cyan/$256$QAM}{75}{160}\\
\hline 
\rotatebox{90}{\hspace{.65cm} GB4 Network}
&\wheelchart{227/green/QSPK, 58/blue/$16$QAM}{60}{38}
&\wheelchart{164/green/QSPK, 75/blue/$16$QAM, 46/red/$64$QAM}{50}{40}
&\wheelchart{164/green/QSPK, 75/blue/$16$QAM, 32/red/$64$QAM, 14/cyan/$256$QAM}{105}{40}\\
\end{tabular}
}
\end{center}
\caption{Percentage of transceivers using different modulation formats for the optimal HD-FEC solution (for \ORVM) and the three network topologies. The results are shown for different maximum constellation sizes $\hat{M}$.}
\label{ModulationFormatsHD}
\end{figure*}

%% file: ModulationFormatsSD.tex
\begin{figure*}
\begin{center}
\setlength{\tabcolsep}{3pt}
\renewcommand{\arraystretch}{1.4}
{\footnotesize
\begin{tabular}{cccc}
\hline 

\hline
Mod. Size & $\hat{M}=16$ & $\hat{M}=64$ & $\hat{M}=256$\\
\hline 

\hline
\rotatebox{90}{\hspace{.65cm} DGT Network}
&\wheelchart{9/green/QSPK, 606/blue/$16$QAM}{70}{258}
&\wheelchart{9/blue/$16$QAM, 606/red/$64$QAM}{70}{343}
&\wheelchart{316/red/$64$QAM, 299/cyan/$256$QAM}{45}{402}\\
\hline 
\rotatebox{90}{\hspace{.65cm} NSF Network}
&\wheelchart{47/green/QSPK, 500/blue/$16$QAM}{69}{146}
&\wheelchart{2/green/QSPK, 379/blue/$16$QAM, 166/red/$64$QAM}{63}{198}
&\wheelchart{251/blue/$16$QAM, 235/red/$64$QAM, 61/cyan/$256$QAM}{110}{200}\\
\hline 
\rotatebox{90}{\hspace{.65cm} GB4 Network}
&\wheelchart{219/green/QSPK, 66/blue/$16$QAM}{60}{46}
&\wheelchart{159/green/QSPK, 77/blue/$16$QAM, 49/red/$64$QAM}{50}{51}
&\wheelchart{59/green/QSPK, 157/blue/$16$QAM, 20/red/$64$QAM, 49/cyan/$256$QAM}{60}{52}\\
\hline
\end{tabular}
}
\end{center}
\caption{Percentage of transceivers using different modulation formats for the optimal SD-FEC solution (for \ORVM) and the three network topologies. The results are shown for different maximum constellation sizes $\hat{M}$.}
\label{ModulationFormatsSD}
\end{figure*}

%% file: ORVarMHDandSD.tex

\pgfplotsset{
   /pgfplots/bar  cycle  list/.style={/pgfplots/cycle  list={%
        {green,fill=green},%
        {blue,fill=blue},%
        {red,fill=red},%
        {cyan,fill=cyan},%
        {purple,fill=purple},%
        {yellow,fill=yellow}
     }
   },
}
\begin{tikzpicture}
\begin{axis}[
	ybar stacked,width=\columnwidth,
	height=0.44\columnwidth,
	bar width=9pt,
	ymin=0,
	ymax=450,
    	xlabel={Maximum Constellation Size $\hat{M}$},
    	xtick={1,2,...,6},
        every axis/.append style={font=\footnotesize},
    	xlabel style={yshift=0.2cm},
    	ytick={0,100,...,450},
    	ylabel={Throughput $\Theta$~[Tbps]},
    	ylabel style={yshift=-0.1cm},
	title=DTG Network,
	title style={yshift=-0.2cm}]
\foreach \i in {1,2,...,6}
\addplot+[ybar,bar shift=-0.20cm] file {OptimumModDist_ORVMHD_DT2015_m_\i.dat};
\end{axis}
\begin{axis}[ticks=none,
	ybar stacked,width=\columnwidth,
	height=0.44\columnwidth,
	bar width=9pt,
	ymin=0,
	ymax=450]
\foreach \i in {1,2,...,6}
\addplot+[ybar,bar shift=0.20cm] file {OptimumModDist_ORVMSD_DT2015_m_\i.dat};
\end{axis}
\begin{axis}[ticks=none,
	width=\columnwidth,
	height=0.44\columnwidth,
	bar width=9pt,
	ymin=0,
	ymax=450]
\addplot[color=red,mark=square*, thick,mark options={fill=white},mark size=1.5pt,xshift=-0.20cm] file {Throughputs_ORVMHD_DT2015.dat};
\end{axis}
\begin{axis}[ticks=none,
	width=\columnwidth,
	height=0.44\columnwidth,
	bar width=9pt,
	ymin=0,
	ymax=450]
\addplot[color=blue,mark=square*, thick,mark options={fill=white},mark size=1.5pt,xshift=0.20cm] file {Throughputs_ORVMSD_DT2015.dat};
\end{axis}
\end{tikzpicture}
\begin{tikzpicture}
\begin{axis}[
	ybar stacked,width=\columnwidth,
	height=0.44\columnwidth,
	bar width=9pt,
	ymin=0,
	ymax=250,
    	xlabel={Maximum Constellation Size $\hat{M}$},
	xtick={1,2,...,6},
        every axis/.append style={font=\footnotesize},
    	xlabel style={yshift=0.2cm},
    	ytick={0,100,...,250},
    	ylabel={Throughput $\Theta$~[Tbps]},
    	ylabel style={yshift=-0.1cm},
	title=NSF Network,
	title style={yshift=-0.2cm}]
\foreach \i in {1,2,...,6}
\addplot+[ybar,bar shift=-0.20cm] file {OptimumModDist_ORVMHD_NSFNET_m_\i.dat};
\end{axis}
\begin{axis}[ticks=none,
	ybar stacked,width=\columnwidth,
	height=0.44\columnwidth,
	bar width=9pt,
	ymin=0,
	ymax=250]
\foreach \i in {1,2,...,6}
\addplot+[ybar,bar shift=0.20cm] file {OptimumModDist_ORVMSD_NSFNET_m_\i.dat};
\end{axis}
\begin{axis}[ticks=none,
	width=\columnwidth,
	height=0.44\columnwidth,
	bar width=9pt,
	ymin=0,
	ymax=250]
\addplot[color=red,mark=square*, thick,mark options={fill=white},mark size=1.5pt,xshift=-0.20cm] file {Throughputs_ORVMHD_NSFNET.dat};
\end{axis}
\begin{axis}[ticks=none,
	width=\columnwidth,
	height=0.44\columnwidth,
	bar width=9pt,
	ymin=0,
	ymax=250]
\addplot[color=blue,mark=square*, thick,mark options={fill=white},mark size=1.5pt,xshift=0.20cm] file {Throughputs_ORVMSD_NSFNET.dat};
\end{axis}
\end{tikzpicture}
\begin{tikzpicture}
\begin{axis}[
	ybar stacked,width=\columnwidth,
	height=0.44\columnwidth,
	bar width=9pt,
	ymin=0,
	ymax=70,
    	xlabel={Maximum Constellation Size $\hat{M}$},
    	xtick={1,2,...,6},
	every axis/.append style={font=\footnotesize},
    	xlabel style={yshift=0.2cm},
    	ytick={0,20,...,70},
    	ylabel={Throughput $\Theta$~[Tbps]},
    	ylabel style={yshift=0.03cm},
	title=GB4 Network,
	title style={yshift=-0.2cm}]
\foreach \i in {1,2,...,6}
\addplot+[ybar,bar shift=-0.20cm] file {OptimumModDist_ORVMHD_GoogleB4_m_\i.dat};
\end{axis}
\begin{axis}[ticks=none,
	ybar stacked,width=\columnwidth,
	height=0.44\columnwidth,
	bar width=9pt,
	ymin=0,
	ymax=70]
\foreach \i in {1,2,...,6}
\addplot+[ybar,bar shift=0.20cm] file {OptimumModDist_ORVMSD_GoogleB4_m_\i.dat};
\end{axis}
\begin{axis}[ticks=none,
	width=\columnwidth,
	height=0.44\columnwidth,
	bar width=9pt,
	ymin=0,
	ymax=70]
\addplot[color=red,mark=square*, thick,mark options={fill=white},mark size=1.5pt,xshift=-0.20cm] file {Throughputs_ORVMHD_GoogleB4.dat};
\end{axis}
\begin{axis}[ticks=none,
	width=\columnwidth,
	height=0.44\columnwidth,
	bar width=9pt,
	ymin=0,
	ymax=70]
\addplot[color=blue,mark=square*, thick,mark options={fill=white},mark size=1.5pt,xshift=0.20cm] file {Throughputs_ORVMSD_GoogleB4.dat};
\end{axis}
\end{tikzpicture}